\documentclass[12pt]{article}
\usepackage{amsmath, amssymb, mathrsfs, bm, natbib}
\usepackage[usenames]{color}
\usepackage[dvips]{graphicx}
\usepackage{rotating}

\usepackage{setspace}
\usepackage{mathrsfs}
\usepackage{amssymb,amsmath,amsthm,mathrsfs,bm,verbatim}
\usepackage{indentfirst}
\usepackage{listings}
\usepackage{booktabs,multirow}

\newtheorem{theorem}{Theorem}

\newcommand{\mbs}[1]{\boldsymbol{#1}}

\numberwithin{equation}{section}
\begin{document}

\begin{center}
\Large{Nonparametric Adaptive CUSUM Chart for Detecting Arbitrary Distributional Changes}
\end{center}

\begin{center}
{\large Jun Li}\\
     Department of Statistics, University of California - Riverside
\end{center}

\begin{abstract}

\noindent
Nonparametric control charts that can detect arbitrary distributional changes are highly desirable due to their flexibility to adapt to different distributional assumptions and distributional changes. However, most of such control charts in the literature either involve some tuning parameter, which needs to be pre-specified, or involve intensive computation. In this paper, we propose a new nonparametric adaptive CUSUM chart for detecting arbitrary distributional changes. The proposed control chart does not depend on any tuning parameter and is efficient in computation. Its self-starting nature makes the proposed control chart applicable to situations where no sufficiently large reference data are available. Our proposed control chart also has a built-in post-signal diagnostics function that can identify what kind of distributional changes have occurred after an alarm. Our simulation study and real data analysis show that the proposed control chart performs well across a broad range of settings, and compares favorably with existing nonparametric control charts.

\end{abstract}


\section{Introduction}\label{sec:intro}

Statistical process control (SPC) applies statistical methods to the monitoring and control of a process in order to detect abnormal variations of the process. One of the most popular SPC tools is the control chart, which plots a statistic that measures a feature of the process over time. When the charting statistic is well within the predetermined control limits, it indicates that the process is in a state of statistical control (hereafter in-control). When this charting statistic goes beyond the control limits, it triggers an alarm to indicate that the process is likely experiencing abnormal variations (hereafter out-of-control). Control charts are easy to visualize and interpret, therefore they have been successfully applied to applications across many different industries, including fraud detection, disease outbreak surveillance, network traffic monitoring and others (see, for example, Tsung et al. (2007), Woodall (2006), Jeske et al. (2009)).

In the SPC literature, there exist parametric control charts and nonparametric control charts. Parametric control charts need to assume a particular parametric distribution for the process. In practice it is often not easy to identify the parametric distribution that would be appropriate for a specific application. If the distribution is not specified correctly, parametric control charts may not perform as expected. In contrast, nonparametric control charts do not require specifying a particular parametric distribution for the process and remain valid regardless of the true underlying distribution. Therefore, nonparametric control charts are more desirable in many real world applications.

There are many nonparametric control charts in the literature. We refer to Chakraborti, van der Laan and Bakir (2001) and Chapter 8 of Qiu (2014) for an overview on this topic. Most of the existing nonparametric control charts were developed to detect location changes only. However, in practical situations it is usually unknown in advance what kind of changes the process will experience. Therefore, it is more desirable to develop a nonparametric control chart that can detect any arbitrary distributional changes. For this purpose, Zou and Tsung (2010) proposed an EWMA chart based on a powerful goodness-of-fit test. However, according to the simulation studies conducted in Ross and Adams (2012), this EWMA chart is only sensitive in detecting scale increases and is not as powerful as its competitors in detecting other types of distributional changes including location shifts. In addition, their proposed EWMA chart involves a weight parameter $\lambda$, which practitioners need to pre-specify. Different choices of $\lambda$ will affect the detection power of the resulting control chart. In general, the EWMA chart with smaller $\lambda$ is more powerful for detecting smaller changes, and the one with larger $\lambda$ is more powerful for detecting larger changes. However, in practice, it is rarely known in advance what kind of changes will occur.

To overcome the above limitations, Ross and Adams (2012) proposed two control charts based on the change-point detection (CPD) framework. Their proposed CPD charts are free of any tuning parameter and are shown to have better overall performance than Zou and Tsung's EWMA chart for detecting different distributional changes. However, like most CPD charts, the computation of their proposed charts is very intensive, since at each time point all the possible change-point scenarios need to be considered.

To detect any arbitrary distributional changes, Qiu and Li (2011) also proposed two nonparametric control charts by first converting the nonparametric problem into a categorical data analysis problem through data categorization and then developing CUSUM charts for monitoring the resulting categorical data. The idea of developing nonparametric control charts through data categorization is very innovative, since it allows adoption of many existing categorical data analysis methods to develop new nonparametric tools in the SPC field. However, similar to the above Zou and Tsung's EWMA chart, the two CUSUM charts proposed by Qiu and Li (2011) involve a tuning parameter $k$, which needs to be pre-specified. In the parametric setting, the optimal choice of $k$ in the CUSUM statistic is usually linked to the out-of-control distribution, therefore practitioners have some general guideline on how to choose $k$. Unfortunately, in the nonparametric CUSUM statistics proposed in Qiu and Li (2011), it is not clear how $k$ is linked to the out-of-control distribution. Because of this, it is not even clear what the right range is for the value of $k$. In their paper, they considered $k=0.001$, 0.005, 0.01, or 0.05, which seems to be much smaller than those commonly used in other CUSUM statistics. According to some simulation study we conducted, the in-control run lengths of their CUSUM statistics with those small values of $k$ have much larger variability than what we usually expect from regular CUSUM statistics. It seems that some larger values of $k$ should be used instead. But again it is not clear what is the right choice of $k$ people should use in practice. Furthermore, based on our simulation studies, the control charts directly based on the  categorial data after data categorization are usually less efficient than other rank-based nonparametric control charts due to the loss of the ordering information from the original data.

To address all the above limitations, in this paper we propose a new nonparametric control chart for detecting arbitrary distributional changes. More specifically, we first follow the above data categorization idea to develop a new CUSUM chart for monitoring the resulting categorical data. The CUSUM chart we propose is more efficient than the ones used in Qiu and Li (2011) for detecting different distributional changes, since it is capable of incorporating the ordering information of the original data. To implement the new CUSUM chart, we need to specify the out-of-control distribution, which is rarely known in advance in practice. To overcome this difficulty, we borrow the idea proposed in Lorden and Pollak (2008) and develop an adaptive version of the proposed CUSUM chart. Our adaptive CUSUM chart does not require the specification of the out-of-control distribution. Instead, it uses the most recent data to estimate the out-of-control distribution. The resulting adaptive CUSUM chart has simple recursive formulas, so it is very efficient in computation and its implementation is simple and straightforward. To address the situation where there are no sufficiently large reference data available, we also develop a self-starting monitoring scheme of the proposed adaptive CUSUM chart. Our simulation studies show that the proposed self-starting adaptive CUSUM chart has better overall performance than other competitors for detecting different distributional changes.

The rest of the paper is organized as follows. In Section 2, we describe our proposed nonparametric adaptive CUSUM chart and its properties. A simulation study is reported in Section 3 to evaluate the performance of our proposed control chart. In Section 4, we demonstrate the application of our proposed control chart using a real data set from a manufacturing process. Finally, we provide some concluding remarks in Section 5. All the proofs are deferred to the Appendix.

\section{Methodology}
\subsection{The proposed CUSUM statistic}\label{sec:CUSUM}
The typical setup we consider in this paper is the following. There are $m$ independent and identically distributed reference (historical) data, denoted by $X_{-m+1}$, ..., $X_{0}$, from some in-control distribution $f_{0,X}$. Let $X_1, X_2, \cdots$ be the future observations collected over time from the process. At any time $t$, we observe $X_1, X_2, \cdots, X_t$, and  the task of control charts at this time $t$ is to decide whether the process has changed based on $X_1, X_2, \cdots, X_t$. This can be formulated as the following hypothesis testing problem,
\[
H_0: X_{1}, \cdots, X_t \text{ follows } f_{0,X},
\]
versus
\begin{equation}
\label{test0}
H_1: \exists \, \tau \in [1,t] \text{ such that } X_1, \cdots, X_{\tau-1} \text{ follows } f_{0,X} \text{ and } X_{\tau}, \cdots, X_t \text{ follows } f_{1,X},
\end{equation}
where $\tau$ is the change point, $f_{1,X} \neq f_{0,X}$ and $f_{1,X}$ is usually referred to as the out-of-control distribution.

If we further assume that $f_{0,X}$ and $f_{1,X}$ are both completely known, to test the hypothesis in (\ref{test0}), the test statistic based on the likelihood ratio method is
\[
S_{t}=\max(0,\max_{1 \leq \tau \leq t} \sum_{i=\tau}^t\log\left\{\frac{f_{1,X}(X_i)}{f_{0,X}(X_i)}\right\}),
\]
and it has the following convenient recursive representation
\begin{equation}
\label{eqn:CUSUM}
S_{t}=\max(0,S_{t-1}+\log\left\{\frac{f_{1,X}(X_t)}{f_{0,X}(X_t)}\right\}).
\end{equation}
The popular CUSUM chart discussed in Page (1954) is then constructed by monitoring the above $S_{t}$ over the time and it raises an alarm if $S_{t}$ exceeds some threshold. The above CUSUM chart is easy to construct and enjoys some optimality property (Moustakides (1986)), therefore it has been widely used in many applications.

To implement the above CUSUM chart, both the in-control and out-of-control distributions, $f_{0,X}$ and $f_{1,X}$, need to be completely specified. However, in our nonparametric setting, both $f_{0,X}$ and $f_{1,X}$ of $X_1,...,X_t$ are unknown. To overcome this difficulty, we first use the data categorization idea introduced in Qiu and Li (2011) to categorize the data so that the in-control and out-of-control distributions of the resulting categorical data can be easily established. More specifically, let $-\infty<q^{(1)}_1<q^{(1)}_2<\cdots<q^{(1)}_{d-1}<\infty$ be the $d-1$ boundary points, and the real line  is then partitioned into the following $d$ intervals,
\[
A^{(1)}_1=(-\infty, q^{(1)}_1],\, A^{(1)}_2=(q^{(1)}_1,q^{(1)}_2],\,...,\, A^{(1)}_d=(q^{(1)}_{d-1},\infty).
\]
Define
\[
Y^{(1)}_{t,j}= I(X_t \in A^{(1)}_j), \quad \text{for } j=1,...,d,
\]
where $I(u)$ is the indicator function that equals 1 when $u$ is true and 0 otherwise. Then $Y^{(1)}_{t,j}$ indicates whether $X_t$ falls in the $j$-th interval $A^{(1)}_j$. Define $\mbs{Y}^{(1)}_t=(Y^{(1)}_{t,1},...,Y^{(1)}_{t,d})'$. It is easy to see that $\mbs{Y}^{(1)}_t$ follows a multinomial distribution with $n=1$ and $p^{(1)}_j=P(X_t \in A^{(1)}_j)$, $j=1,...,d$, denoted by Multi$(1;p^{(1)}_1,...,p^{(1)}_d)$. Therefore, based on the above data categorization, the original data $X_t$ with any arbitrary distribution is converted into the multinomial random variable $\mbs{Y}^{(1)}_t$.

To completely characterize the distribution of $\mbs{Y}^{(1)}_t$, we need to know $\{q^{(1)}_1,q^{(1)}_2,...,q^{(1)}_{d-1}\}$. Following Qiu and Li (2011), we choose $q^{(1)}_j$ to be the $(j/d)$-th quantile of the in-control distribution of $X_t$. Then the in-control distribution $f_{0,Y^{(1)}}$ of the $\mbs{Y}^{(1)}_t$ is simply Multi$(1;1/d,...,1/d)$. Based on those $q^{(1)}_j$'s, we first assume that the out-of-control distribution $f_{1,Y^{(1)}}$ of $\mbs{Y}^{(1)}_t$ is given by another multinomial distribution Multi$(1; p^{(1)}_1,...,p^{(1)}_d)$, where $\sum_{j=1}^dp^{(1)}_j=1$ and $(p^{(1)}_1,...,p^{(1)}_d)\neq (1/d,...,1/d)$. Using the in-control and out-of-control distributions of $\mbs{Y}^{(1)}_t$ instead of those of $X_t$, the CUSUM statistic in (\ref{eqn:CUSUM}) becomes
\begin{align}
\label{eqn:CUSUM0}
S^{(0)}_{t}&=\max\Big(0,S^{(0)}_{t-1}+\log\left\{\frac{f_{1,Y^{(1)}}(\mbs{Y}^{(1)}_t)}{f_{0,Y^{(1)}}(\mbs{Y}^{(1)}_t)}\right\}\Big) \nonumber\\
     &=\max\Big(0,S^{(0)}_{t-1}+ \sum_{j=1}^d Y^{(1)}_{t,j}\log(dp^{(1)}_j)\Big).
\end{align}

Similar to the charting statistics proposed in Qiu and Li (2011), the above CUSUM statistic is usually less powerful than other rank-based charting statistics. The reason is that the ordering information of the original data $X_t$ is lost in (\ref{eqn:CUSUM0}), since it does not make use of the ordering information of the $d$ intervals, $A^{(1)}_1$,..., $A^{(1)}_d$.
To overcome this drawback, we need to find a new way to construct the CUSUM statistic so that the ordering information of $A^{(1)}_1$,..., $A^{(1)}_d$ can be used. For this purpose, we first define the cumulative unions of $A^{(1)}_1$,..., $A^{(1)}_d$, i.e.,
\[
A^{(1)}_1, \quad A^{(1)}_1 \cup A^{(1)}_2, \quad A^{(1)}_1 \cup A^{(1)}_2 \cup A^{(1)}_3, \quad ..., \quad A^{(1)}_1 \cup \cdots \cup A^{(1)}_d.
\]
Similarly we define the cumulative sums of $Y^{(1)}_{t,1},...,Y^{(1)}_{t,d}$, i.e.,
\[
Z^{(1)}_{t,j}=\sum_{l=1}^j Y^{(1)}_{t,l}, \quad j=1,...,d.
\]
Then $Z^{(1)}_{t,j}$ indicates whether $X_t$ falls in the interval $A^{(1)}_1 \cup \cdots \cup A^{(1)}_j$. Write $\mbs{Z}^{(1)}_t=(Z^{(1)}_{t,1},...,Z^{(1)}_{t,d})'$. The new vector $\mbs{Z}^{(1)}_t$ contains the same amount of information as $\mbs{Y}^{(1)}_t$. However, if we use the log-likelihood ratio based on $\mbs{Z}^{(1)}_t$ in our CUSUM statistic, the ordering information of $A^{(1)}_1$,..., $A^{(1)}_d$ can be incorporated, so the ordering information of $X_t$ can be preserved.

To develop the log-likelihood ratio based on $\mbs{Z}^{(1)}_t$, we first notice that $Z^{(1)}_{t,j}$, $j=1,...,d-1$, is a Bernoulli random variable and the log-likelihood ratio based on $Z^{(1)}_{t,j}$ is
\[
Z^{(1)}_{t,j}\log\left(\frac{\sum_{l=1}^j p^{(1)}_l}{j/d}\right)+(1-Z^{(1)}_{t,j})\log\left(\frac{1-\sum_{l=1}^j p^{(1)}_l}{1-j/d}\right).
\]
Then our proposed log-likelihood ratio based on $\mbs{Z}^{(1)}_t$ is simply the weighted sum of the above log-likelihood ratios, i.e.,
\[
\log\left\{\frac{f_{1,Z^{(1)}}(\mbs{Z}^{(1)}_t)}{f_{0,Z^{(1)}}(\mbs{Z}^{(1)}_t)}\right\}=\sum_{j=1}^{d-1} \omega(j)\Big\{ Z^{(1)}_{t,j}\log\left(\frac{\sum_{l=1}^j p^{(1)}_l}{j/d}\right)+(1-Z^{(1)}_{t,j})\log\left(\frac{1-\sum_{l=1}^j p^{(1)}_l}{1-j/d}\right) \Big\},
\]
where $\omega(j)$ is the weight function, and we choose $\omega(j)=(j/d)^{-1}(1-j/d)^{-1}$ to give more weights to the tail areas. Therefore, our proposed CUSUM statistic is
\begin{align}
\label{eqn:CUSUM_loc}
S^{(1)}_{t}&=\max\Big(0,S^{(1)}_{t-1}+\log\left\{\frac{f_{1,Z^{(1)}}(\mbs{Z}^{(1)}_t)}{f_{0,Z^{(1)}}(\mbs{Z}^{(1)}_t)}\right\}\Big)\nonumber\\
     &=\max\Big(0,S^{(1)}_{t-1}+ \sum_{j=1}^{d-1} \frac{d^2}{j(d-j)}\Big\{ Z^{(1)}_{t,j}\log\left(\frac{\sum_{l=1}^j p^{(1)}_l}{j/d}\right)+(1-Z^{(1)}_{t,j})\log\left(\frac{1-\sum_{l=1}^j p^{(1)}_l}{1-j/d}\right) \Big\}\Big).
\end{align}

As described above, using the log-likelihood ratio of $\mbs{Z}^{(1)}_t$ in our CUSUM statistic helps preserve the ordering information of the data. Based on how both the $d$ intervals, $A^{(1)}_1$,..., $A^{(1)}_d$, and their cumulative unions are constructed, the ordering information of the data used in the above CUSUM statistic is from the smallest to the largest. In the nonparametric literature, the Wilcoxon-Mann-Whitney test is a powerful test for testing location differences, and the Ansari-Bradley test is a powerful test for testing scale differences. Both tests can be considered as a rank-sum test. In the Wilcoxon-Mann-Whitney test, the data are ranked from the smallest to the largest, while in the  Ansari-Bradley test, the data can be considered as being ranked from the center outward. This observation makes us believe that, although our CUSUM statistic in (\ref{eqn:CUSUM_loc}) can detect any arbitrary distributional changes, it might not be very powerful for detecting scale changes. To develop a CUSUM statistic that is efficient for scale changes, we need to make use of the center-outward ordering of the data.

To do so, different from how we categorize the data previously, we categorize the data in a center-outward fashion. More specifically, let $q^{(2)}_j$, $j=1,...,2d-1$, be the $(j/(2d))$-th quantile of the in-control distribution of $X_t$. We partition the real line into the following $d$ regions,
\begin{align*}
A^{(2)}_1&=(q^{(2)}_{d-1}, q^{(2)}_{d+1}],\\
A^{(2)}_2&=(q^{(2)}_{d-2},q^{(2)}_{d-1}] \bigcup (q^{(2)}_{d+1},q^{(2)}_{d+2}],\\ A^{(2)}_3&=(q^{(2)}_{d-3},q^{(2)}_{d-2}] \bigcup (q^{(2)}_{d+2},q^{(2)}_{d+3}],\\
& \cdots \, \cdots\\
A^{(2)}_d&=(-\infty, q^{(2)}_{1}] \bigcup (q^{(2)}_{2d-1},\infty).
\end{align*}

It is clear that $A^{(2)}_1$,...,$A^{(2)}_d$ are ordered from the center outward. Define $Y^{(2)}_{t,j}= I(X_t \in A^{(2)}_j)$. It is easy to see that $\mbs{Y}^{(2)}_t=(Y^{(2)}_{t,1},...,Y^{(2)}_{t,d})'$ follows a multinomial distribution and its in-control distribution is Multi$(1;1/d,...,1/d)$. Again we assume that the out-of-control distribution of $\mbs{Y}^{(2)}_t$ is given by another multinomial distribution Multi$(1; p^{(2)}_1,...,p^{(2)}_d)$, where $\sum_{j=1}^dp^{(2)}_j=1$ and $(p^{(2)}_1,...,p^{(2)}_d)\neq (1/d,...,1/d)$. Although $A^{(2)}_1$,...,$A^{(2)}_d$ are ordered from the center outward, if we use $\mbs{Y}^{(2)}_t$ directly to construct the CUSUM statistic, the center-outward ordering of $A^{(2)}_1$,...,$A^{(2)}_d$ will not be utilized. Similar to how we construct $S^{(1)}_{t}$ in (\ref{eqn:CUSUM_loc}) to incorporate the left-to-right ordering information of the data, we consider the cumulative unions of $A^{(2)}_1$,...,$A^{(2)}_d$,
\[
A^{(2)}_1, \quad A^{(2)}_1 \cup A^{(2)}_2, \quad A^{(2)}_1 \cup A^{(2)}_2 \cup A^{(2)}_3, \quad ..., \quad A^{(2)}_1 \cup \cdots \cup A^{(2)}_d,
\]
and the cumulative sums of $Y^{(2)}_{t,1},...,Y^{(2)}_{t,d}$,
\[
Z^{(2)}_{t,j}=\sum_{l=1}^j Y^{(2)}_{t,l}, \quad j=1,...,d.
\]
Using the same method for obtaining $S^{(1)}_{t}$ in (\ref{eqn:CUSUM_loc}), we can obtain the following CUSUM statistic that makes use of the center-outward ordering information of the data,
\begin{align}
\label{eqn:CUSUM_scal}
S^{(2)}_{t}&=\max\Big(0,S^{(2)}_{t-1}+\log\left\{\frac{f_{1,Z^{(2)}}(\mbs{Z}^{(2)}_t)}{f_{0,Z^{(2)}}(\mbs{Z}^{(2)}_t)}\right\}\Big)\nonumber\\
     &=\max\Big(0,S^{(2)}_{t-1}+ \sum_{j=1}^{d-1} \frac{d^2}{j(d-j)}\Big\{ Z^{(2)}_{t,j}\log\left(\frac{\sum_{l=1}^j p^{(2)}_l}{j/d}\right)+(1-Z^{(2)}_{t,j})\log\left(\frac{1-\sum_{l=1}^j p^{(2)}_l}{1-j/d}\right) \Big\}\Big).
\end{align}

Both $S^{(1)}_t$ and $S^{(2)}_t$ can be used to detect any arbitrary distributional changes. As shown in our simulation study in Section 3.2, $S^{(1)}_t$ is more powerful than $S^{(2)}_t$ for detecting location changes, since it uses the left-to-right ordering information of the data. In contrast, $S^{(2)}_t$ uses the center-outward ordering information of the data, therefore it is more powerful than $S^{(1)}_t$ for detecting scale changes. If no prior information is available on what type of changes the process might experience, we propose to use the following CUSUM statistic,
\begin{equation}
\label{eqn:CUSUM_loc_scal}
S_t=\max(S^{(1)}_t,S^{(2)}_t).
\end{equation}

\subsection{The adaptive CUSUM statistic}
\label{sec:adaptiveCUSUM}
To implement the above CUSUM statistic $S_t$,  $\{p^{(1)}_1,...,p^{(1)}_d\}$ and $\{p^{(2)}_1,...,p^{(2)}_d\}$ in the out-of-control distributions of $\mbs{Y}^{(1)}_t$ and  $\mbs{Y}^{(2)}_t$ need to be specified in advance. This can be a difficult task for many real-world applications, where prior knowledge of the out--of-control distribution may not be available. This is the case even for the standard CUSUM statistic when both the in-control and out-of-control distributions are the normal distributions but with different means. To circumvent this difficulty, a few adaptive CUSUM statistics were proposed in the literature. For example, in Sparks (2000), instead of using the specified out-of-control mean in the standard CUSUM statistic, an estimate of the out-of-control mean using an exponentially weighted moving average of all the past observations is plugged in. In Han and Tsung (2006), the absolute value of the current observation is used as the estimate of the out-of-control mean in the standard CUSUM statistic. Following the same idea, Lorden and Pollak (2008) proposed another way to estimate the out-of-control mean to be used in the CUSUM statistic, and proved the asymptotic optimality of the resulting CUSUM statistic under a single-parameter exponential family. Recently, Wu (2016) generalized Lorden and Pollak's result to the multi-parameter exponential family. In both Lorden and Pollak (2008) and Wu(2016), the key observation is that, at any given time $t$, the most recent time $\hat{\tau}$ when the CUSUM statistic goes back to 0 provides a candidate estimate for the possible change point $\tau$, and therefore the observations collected after $\hat{\tau}$ can be used to estimate the parameters in the out-of-control distribution.

In the following, we adopt the approach from Lorden and Pollak (2008) and Wu (2016) and substitute $\{p^{(i)}_1,...,p^{(i)}_d\}$ ($i=1,2$) in our proposed CUSUM statistic $S_t$ by their estimates based on the observations collected after their change point estimates $\hat{\tau}^{(i)}$, where $\hat{\tau}^{(i)}$ is the most recent time when the CUSUM statistic $S^{(i)}_t$ equals 0. More specifically, define, for $i=1,2$, $t \geq 1$,
\begin{equation}
\label{eqn:adaptiveCUSUM}
\hat{S}^{(i)}_{t}=\max\Big(0,\hat{S}^{(i)}_{t-1}+ \sum_{j=1}^{d-1} \frac{d^2}{j(d-j)}\Big\{ Z^{(i)}_{t,j}\log\left(\frac{\sum_{l=1}^j \hat{p}^{(i)}_{t,l}}{j/d}\right)+(1-Z^{(i)}_{t,j})\log\left(\frac{1-\sum_{l=1}^j \hat{p}^{(i)}_{t,l}}{1-j/d}\right) \Big\}\Big),
\end{equation}
where the $\hat{p}^{(i)}_{t,l}$ are the estimates of the $p^{(i)}_l$ at time $t$ and are defined by
\begin{equation}
\label{eqn:phat}
\hat{p}^{(i)}_{t,l}=\frac{\alpha_l+N^{(i)}_{t,l}}{\sum_{j=1}^d \alpha_j+ N^{(i)}_t}.
\end{equation}
In the above estimates,  $N^{(i)}_t$ is the number of observations collected before the current time $t$ but after the candidate change point estimate $\hat{\tau}^{(i)}$. Similarly, $N^{(i)}_{t,l}$ is the number of observations falling in the $l$th interval $A^{(i)}_l$ before time $t$ but after time $\hat{\tau}^{(i)}$.  Both $N^{(i)}_t$ and $N^{(i)}_{t,l}$ can be calculated recursively by
\begin{align*}
N^{(i)}_{t}&=\begin{cases}
N^{(i)}_{t-1}+1, & \text{if } \hat{S}^{(i)}_{t-1}>0, \\
0, &  \text{if } \hat{S}^{(i)}_{t-1}=0,
\end{cases}\\
N^{(i)}_{t,l}&=
\begin{cases}
N^{(i)}_{t-1,l}+ Y^{(i)}_{t-1,l}, & \text{if } \hat{S}^{(i)}_{t-1}>0, \\
0, &  \text{if } \hat{S}^{(i)}_{t-1}=0.
\end{cases}
\end{align*}

The constants $\{\alpha_1,...,\alpha_d\}$ in (\ref{eqn:phat}) can be considered as the parameters of the Dirichlet distribution, the conjugate prior for $\{p^{(i)}_1,...,p^{(i)}_d\}$. Therefore, the above estimate $\hat{p}^{(i)}_{t,l}$ can be considered as a Bayesian estimate. In Bayesian statistics, it is common to choose  $\alpha_1=\cdots=\alpha_d=1$ as the noninformative prior for $\{p^{(i)}_1,...,p^{(i)}_d\}$. However, in our case a closer examination of $\hat{p}_{t,l}$ reveals that, whenever $\hat{S}^{(i)}_{t}$ returns to 0, $\alpha_l/\sum_{j=1}^d\alpha_j$ will be used to estimate $p^{(i)}_{l}$. Therefore,  the choice $\alpha_1=\cdots=\alpha_d=1$ does not work. Instead, we can choose $\{\alpha_1,...,\alpha_d\}$ proportional to $\{p^{(i)}_1,...,p^{(i)}_d\}$ when the process experiences the smallest distributional change that is meaningful. In this paper, we choose $\{\alpha_1,...,\alpha_d\}$ as follows. We first assume that the in-control distribution of $X_t$ is $N(0,1)$ and its smallest meaningful out-of-control distribution is either $N(0.25,1)$ or $N(-0.25,1)$. Under this in-control and out-of-control distributional assumption for $X_t$, we can obtain the corresponding out-of-control distribution of $\mbs{Y}^{(1)}_t$, denoted by Multi$(1; p^+_{1},...,p^+_{d})$ for $N(0.25,1)$ and Multi$(1; p^-_{1},...,p^-_{d})$ for $N(-0.25,1)$. Then we choose $\alpha_j=dp^+_{j}$ or $dp^-_{j}$,  $j=1,...,d$. When using $\alpha_{j}=dp^+_{j}$ in $\hat{S}^{(1)}_t$, denoted by $\hat{S}^{(1+)}_t$, the prior indicates a positive location shift, so $\hat{S}^{(1+)}_t$ is more powerful for detecting positive location shifts. When using $\alpha_{j}=dp^-_{j}$ in $\hat{S}^{(1)}_t$, denoted by $\hat{S}^{(1-)}_t$,  the prior indicates a negative location shift, so $\hat{S}^{(1-)}_t$ is more powerful for detecting negative location shifts. Similarly, when using $\alpha_{j}=dp^+_{j}$ in $\hat{S}^{(2)}_t$, denoted by $\hat{S}^{(2+)}_t$, the prior indicates a scale increase, so $\hat{S}^{(2+)}_t$ is more powerful for detecting scale increases. When using $\alpha_{j}=dp^-_{j}$ in $\hat{S}^{(2)}_t$, denoted by $\hat{S}^{(2-)}_t$,  the prior indicates a scale decrease, so $\hat{S}^{(2-)}_t$ is more powerful for detecting scale decreases. If we do not have any prior information about what type of changes the process might encounter, the charting statistic we use is
\begin{equation}
\label{eqn:adaptiveCUSUM_loc_scal}
\hat{S}_t=\max(\hat{S}^{(1+)}_t,\hat{S}^{(1-)}_t,\hat{S}^{(2+)}_t,\hat{S}^{(2-)}_t),
\end{equation}
which is efficient to detect any type of distributional changes.

\subsection{Determining the control limit}\label{sec:controllimit}
\label{sec:controllimit}
As described in the previous section, our proposed adaptive CUSUM statistic is simply $\hat{S}_t=\max(\hat{S}^{(1+)}_t,\hat{S}^{(1-)}_t,\hat{S}^{(2+)}_t,\hat{S}^{(2-)}_t)$, and the resulting control chart is to monitor $\hat{S}_t$ over time $t$, and it raises an alarm if $\hat{S}_t$ exceeds the control limit $h$. As we can see from (\ref{eqn:adaptiveCUSUM}), $\hat{S}_t$ is a function of $\mbs{Y}^{(1)}_t$ and  $\mbs{Y}^{(2)}_t$ only.
Define
\begin{align*}
&B^{(1)}_1=\left(0,\frac{1}{d}\right],\, B^{(1)}_2=\left(\frac{1}{d},\frac{2}{d}\right],\,...,\, B^{(1)}_d=\left(\frac{d-1}{d},1\right),\\
&B^{(2)}_1=\left(\frac{d-1}{2d},\frac{d+1}{2d}\right], \, B^{(2)}_2=\left(\frac{d-2}{2d},\frac{d-1}{2d}\right] \bigcup \left(\frac{d+1}{2d},\frac{d+2}{2d}\right], \, ...,\, B^{(2)}_d=\left(0,\frac{1}{2d}\right] \bigcup \left(\frac{2d-1}{2d},1\right),
\end{align*}
and for $i=1, 2$,
\[
U^{(i)}_{j}= I(U \in B^{(i)}_j), \quad \text{for } j=1,...,d,
\]
where $U$ is a uniform random variable on (0,1). Let $\mbs{U}^{(1)}=(U^{(1)}_1,...,U^{(1)}_d)'$ and $\mbs{U}^{(2)}=(U^{(2)}_1,...,U^{(2)}_d)'$. Then based on the probability integral transformation, it is easy to see that the in-control joint distribution of $\mbs{Y}^{(1)}_t$ and  $\mbs{Y}^{(2)}_t$ is the same as the joint distribution of $\mbs{U}^{(1)}$ and  $\mbs{U}^{(2)}$. Therefore, our proposed adaptive CUSUM control chart based on $\hat{S}_t$ is distribution-free. Determining the control limit $h$ for this CUSUM chart can be achieved by simulating data from any standard continuous distribution, say the standard normal distribution, as $X_t$ and finding $h$ to obtain the desired in-control average run length (denoted by $ARL_0$) through a bi-section search. Table \ref{tab:h} shows the computed control limit $h$ using the bi-section search algorithm based on 10,000 replications for different choices of $d$ when $ARL_0=200, 370, 500, 1000$.

\begin{table}[ht]
\centering
\caption{The computed control limit $h$ for our proposed adaptive CUSUM chart based on 10,000 replications for different choice of $d$ when $ARL_0=200, 370, 500, 1000$.\label{tab:h}}
\begin{tabular}{r|cccc}
  \hline
$ARL_0$ & $d=10$ & $d=20$ & $d=30$ & $d=40$ \\
  \hline
200 & 90.275 & 185.466 & 281.644 & 379.191 \\
370 & 105.941 & 218.886 & 333.933 & 449.201 \\
500 & 113.308 & 235.241 & 358.960 & 483.987 \\
1000& 131.299 & 273.411 & 418.364 & 564.137 \\
   \hline
\end{tabular}
\end{table}

\subsection{Self-starting monitoring scheme}\label{sec:sefstarting}
\label{sec:selfstart}
To categorize the original data $X_t$ and implement our proposed control chart based on $\hat{S}_t$, we need to know $\{q^{(1)}_j\}_{j=1}^{d-1}$ and $\{q^{(2)}_j\}_{j=1}^{2d-1}$, which are the $(j/d)$-th quantile and $(j/(2d))$-th quantile of the in-control distribution of $X_t$, respectively. Since those quantiles are rarely known in practice, we can approximate them by their sample estimates from the in-control reference data. However, in order for the effect of using those quantile estimates instead of the true values on the $ARL_0$ to be negligible,  it usually requires a substantial amount of  in-control reference data. In many real-world applications, it can be very challenging to have such data. To solve this problem, we develop a self-starting monitoring scheme where the estimates of quantiles $\{q^{(1)}_j\}_{j=1}^{d-1}$ and $\{q^{(2)}_j\}_{j=1}^{2d-1}$ are updated sequentially each time when a new observation is collected.

More specifically, at time $t$ we have $m+t-1$ observations collected in the past, i.e.,
\[
X_{-m+1},...,X_0, X_1,...,X_{t-1}.
\]
Let $X_{t,(1)}\leq X_{t,(2)} < \cdots < X_{t,(m+t-1)}$ denote their order statistics. For a given $j$, $j=1,...,2d-1$, find the integer $l$ such that $1\leq l \leq m+t-2$ and
\[
\frac{l}{m+t} \leq \frac{j}{2d} \leq \frac{l+1}{m+t}.
\]
Then based on $X_{-m+1},...,X_0, X_1,...,X_{t-1}$, the $(j/(2d))$-th quantile of the in-control distribution of $X_t$,  $q^{(2)}_j$, can be estimated by
\begin{equation}
\label{eqn:quantiles}
\hat{q}^{(2)}_{t,j}= \left(1-\frac{j(m+t)}{2d}+l\right) X_{t,(l)}+\left(\frac{j(m+t)}{2d}-l\right) X_{t,(l+1)}.
\end{equation}
Since $q^{(1)}_j=q^{(2)}_{2j}$ for $j=1,...,d-1$, the estimates of $q^{(1)}_j$ can be obtained accordingly.

Using those estimates, at time $t$ we partition the real line into the following $d$ left-to-right regions,
\[
\hat{A}^{(1)}_{t,1}=(-\infty, \hat{q}^{(1)}_{t,1}],\, \hat{A}^{(1)}_{t,2}=(\hat{q}^{(1)}_{t,1},\hat{q}^{(1)}_{t,2}],\,...,\, \hat{A}^{(1)}_{t,d}=(\hat{q}^{(1)}_{t,d-1},\infty),
\]
or the following $d$ center-outward regions,
\begin{align*}
\hat{A}^{(2)}_{t,1}&=(\hat{q}^{(2)}_{t,d-1}, \hat{q}^{(2)}_{t,d+1}],\\
\hat{A}^{(2)}_{t,2}&=(\hat{q}^{(2)}_{t,d-2},\hat{q}^{(2)}_{t,d-1}] \bigcup (\hat{q}^{(2)}_{t,d+1},\hat{q}^{(2)}_{t,d+2}],\\
& \cdots \, \cdots\\
\hat{A}^{(2)}_{t,d}&=(-\infty, \hat{q}^{(2)}_{t,1}] \bigcup (\hat{q}^{(2)}_{t,2d-1},\infty).
\end{align*}
Define $\hat{\mbs{Y}}^{(1)}_{t}=(\hat{Y}^{(1)}_{t,1},...,\hat{Y}^{(1)}_{t,d})'$ and $\hat{\mbs{Y}}^{(2)}_{t}=(\hat{Y}^{(2)}_{t,1},...,\hat{Y}^{(2)}_{t,d})'$, where
\[
\hat{Y}^{(1)}_{t,j}= I(X_t \in \hat{A}^{(1)}_{t,j}) \quad  \text{ and } \quad \hat{Y}^{(2)}_{t,j}= I(X_t \in \hat{A}^{(2)}_{t,j}), \quad \text{ for } j=1,...,d.
\]
The following result shows the in-control distributions of $\hat{\mbs{Y}}^{(1)}_{t}$ and $\hat{\mbs{Y}}^{(2)}_{t}$.

\begin{theorem}
\label{thm1}
For $i=1$, $2$, $\hat{\mbs{Y}}^{(i)}_{t}$ are independent and identically distributed as Multi$(1;1/d,...,1/d)$ when the process is in-control.
\end{theorem}

Based on the above result, $\hat{\mbs{Y}}^{(i)}_{t}$ has the same in-control distribution as $\mbs{Y}^{(i)}_{t}$, $i=1,2$. Therefore, in our self-starting monitoring scheme, we replace $\mbs{Y}^{(i)}_{t}$ in our proposed adaptive CUSUM statistic described in Section \ref{sec:adaptiveCUSUM} by $\hat{\mbs{Y}}^{(i)}_{t}$, and the resulting self-starting control chart can still use the control limit we obtain from Section \ref{sec:controllimit}.

In the above self-starting monitoring scheme, it is assumed that the calculation of our sequential quantile estimates (\ref{eqn:quantiles}) starts from $t=1$. In order for Theorem 1 to hold, the size of the reference data $m$ is at least $2d-1$, since this ensures that, for any $t\geq 1$ and any $j$, $j=1,...,2d-1$, we can find an integer $l$ such that $1\leq l \leq m+t-2$ and
\[
\frac{l}{m+t} \leq \frac{j}{2d} \leq \frac{l+1}{m+t}.
\]
If the number of observations we have is smaller than $2d-1$, it implies that we can not find such an integer $l$ for some $j$. If this is the case, we simply define
\[
\hat{q}^{(2)}_{t,j}=\begin{cases}
 X_{t,(1)}, \quad & \text{ if } j/2d < 1/(m+t)\\
 X_{t,(m+t-1)}, \quad & \text{ if } j/2d > (m+t-1)/(m+t)\\
 \end{cases}
\]
When using the above $\hat{q}^{(2)}_{t,j}$, the in-control distribution of $\hat{\mbs{Y}}^{(2)}_{t}$ is not exactly Multi$(1;1/d,...,1/d)$.
Therefore, if $m<2d-1$, the in-control distribution of $\hat{\mbs{Y}}^{(2)}_{t}$ is a little off from its expected one for $t<2d-m$. Since this is the case only for $t<2d-m$, we expect that its effect on the $ARL_0$ is negligible if $2d-m$ is not large.

In the following, we report a simulation study to evaluate such effects. In the simulation study, we choose the size of the reference data $m=10$ or $20$ and the number of categories the data are categorized into $d$=10, 20, 30, or 40. Three different in-control distributions, $f_{0,X}$, are considered: the standard normal, denoted by $N(0,1)$; the $t$ distribution with 2.5 degrees of freedom, denoted by $t(2.5)$; the lognormal distribution with parameters $\mu=1$ and $\sigma=0.5$, denoted by $LN(1,0.5)$. Using the control limits reported in Table \ref{tab:h}, we apply our proposed self-starting monitoring scheme to the data simulated from the above three in-control distributions, and record the time it takes to trigger an alarm, which is the in-control run length. This is repeated 10,000 times and the average of the 10,000 in-control run lengths is the simulated $ARL_0$ of our proposed self-starting monitoring scheme. Table \ref{tab:arl0} shows the simulated $ARL_0$ along with their corresponding standard errors (in the parentheses) under different settings.

\begin{table}[!htbp]
\centering
\caption{The simulated $ARL_0$ for our proposed self-starting adaptive CUSUM chart based on 10,000 replications for different choices of $m$ and $k$ when $ARL_0=200, 370, 500, 1000$.\label{tab:arl0}}
\begin{tabular}{|l|l||cccc|}
  \hline
  & &\multicolumn{4}{|c|}{$ARL_0=200$} \\
\cline{3-6}
$m$ & $f_{0,X}$ & $d=10$ & $d=20$ & $d=30$ & $d=40$\\
  \hline
 & $N(0,1)$ & 201.13(1.86) & 196.13(1.80) & 187.86(1.77) & 181.96(1.78) \\
10 & $t(2.5)$ & 200.11(1.87) & 194.62(1.80) & 189.36(1.78) & 180.87(1.77) \\
 & $LN(1,0.5)$ & 201.40(1.86) & 196.27(1.80) & 187.77(1.77) & 181.93(1.78) \\
  \hline
 & $N(0,1)$ & 200.99(1.87) & 199.01(1.80) & 197.10(1.77) & 195.10(1.80) \\
20 & $t(2.5)$ & 201.56(1.89) & 199.98(1.82) & 198.27(1.79) & 194.52(1.79) \\
 & $LN(1,0.5)$ & 200.68(1.86) & 198.85(1.80) & 197.14(1.78) & 195.02(1.80) \\
  \hline
  \hline
  & &\multicolumn{4}{|c|}{$ARL_0=370$} \\
\cline{3-6}
$m$ & $f_{0,X}$ & $d=10$ & $d=20$ & $d=30$ & $d=40$\\
  \hline
 & $N(0,1)$ & 372.60(3.50) & 366.54(3.41) & 361.38(3.43) & 349.79(3.41) \\
10 & $t(2.5)$ & 368.37(3.53) & 367.35(3.43) & 360.51(3.38) & 348.35(3.39) \\
 & $LN(1,0.5)$ & 372.00(3.50) & 365.84(3.41) & 361.63(3.43) & 348.83(3.41) \\
  \hline
 & $N(0,1)$ & 372.14(3.51) & 369.11(3.38) & 373.16(3.44) & 364.74(3.40) \\
20 & $t(2.5)$ & 368.60(3.55) & 371.98(3.46) & 370.59(3.37) & 365.27(3.40) \\
 & $LN(1,0.5)$ & 371.42(3.52) & 368.73(3.38) & 372.99(3.43) & 364.44(3.40) \\
  \hline
  \hline
  & &\multicolumn{4}{|c|}{$ARL_0=500$} \\
\cline{3-6}
$m$ & $f_{0,X}$ & $d=10$ & $d=20$ & $d=30$ & $d=40$\\
  \hline
 & $N(0,1)$ & 499.75(4.74) & 491.82(4.63) & 482.85(4.63) & 478.76(4.65) \\
10 & $t(2.5)$ & 497.66(4.78) & 501.00(4.75) & 494.74(4.72) & 478.32(4.65) \\
 & $LN(1,0.5)$ & 499.81(4.74) & 491.05(4.63) & 482.64(4.64) & 478.17(4.65) \\
  \hline
  & $N(0,1)$ & 499.29(4.75) & 496.14(4.64) & 496.29(4.64) & 498.95(4.65) \\
20 & $t(2.5)$ & 497.02(4.77) & 504.20(4.72) & 507.01(4.72) & 497.49(4.65) \\
 & $LN(1,0.5)$ & 499.48(4.75) & 495.77(4.64) & 495.71(4.65) & 499.33(4.65) \\
  \hline
  \hline
  & &\multicolumn{4}{|c|}{$ARL_0=1000$} \\
\cline{3-6}
$m$ & $f_{0,X}$ & $d=10$ & $d=20$ & $d=30$ & $d=40$\\
  \hline
 & $N(0,1)$ & 990.69(9.47) & 990.39(9.69) & 988.51(9.76) & 965.14(9.64) \\
10 & $t(2.5)$ & 989.89(9.65) & 992.05(9.66) & 991.03(9.71) & 966.82(9.59) \\
 & $LN(1,0.5)$ & 991.64(9.48) & 990.65(9.68) & 988.75(9.76) & 965.21(9.64) \\
  \hline
 & $N(0,1)$ & 989.05(9.47) & 995.48(9.68) & 999.60(9.75) & 982.38(9.58) \\
20 & $t(2.5)$ & 988.25(9.64) & 998.71(9.63) & 1005.20(9.72) & 994.89(9.65) \\
 & $LN(1,0.5)$ & 989.45(9.47) & 995.54(9.68) & 999.78(9.75) & 982.91(9.58) \\
  \hline
\end{tabular}
\end{table}

As mentioned above, only the first $2d-m$ observations can potentially cause the $ARL_0$ to deviate from the nominal level. To make such effects to be negligible, $2d-m$ should not be very large. This implies that the minimal size of the reference data we need to maintain the desired $ARL_0$ should increase as $d$ increases. As we can see from Table \ref{tab:arl0}, for $d=10$ or 20, the simulated $ARL_0$ are  close to the nominal level even when $m=10$. However, for $d=30$ or 40, when $m=10$, the simulated $ARL_0$ can deviate from the nominal level, indicating the size of the reference data $m$ need to increase in those cases. Based on our simulations, $m=20$ seems to work well for all the cases considered here.

\subsection{Post-signal diagnostics}\label{sec:postsignal}
When using the control chart to monitor the process in practice, in addition to detecting a change as quickly as possible, it is also important to identify what kind of distributional changes have triggered the alarm. In the literature, most of the existing nonparametric control charts have to implement extra tests to identify what kind of distributional changes have occurred after an alarm. Different from those methods, our proposed adaptive CUSUM chart can identify the distributional change automatically when the alarm is triggered. To see this, recall that our adaptive CUSUM chart simply monitors $\hat{S}_t=\max(\hat{S}^{(1+)}_t,\hat{S}^{(1-)}_t,\hat{S}^{(2+)}_t,\hat{S}^{(2-)}_t)$, and it raises an alarm whenever $\hat{S}_t$ exceeds some control limit $h$. Because $\hat{S}^{(1+)}_t$, $\hat{S}^{(1-)}_t$, $\hat{S}^{(2+)}_t$, and $\hat{S}^{(2-)}_t$ all have the same in-control distribution of run lengths, our proposed monitoring scheme is equivalent to monitoring $\hat{S}^{(1+)}_t$, $\hat{S}^{(1-)}_t$, $\hat{S}^{(2+)}_t$, and $\hat{S}^{(2-)}_t$ separately, and raising an alarm whenever at least one of them exceeds $h$. Recall that $\hat{S}^{(1+)}_t$ is more powerful for detecting positive location shifts, $\hat{S}^{(1-)}_t$ is more powerful for detecting negative location shifts, $\hat{S}^{(2+)}_t$ is more powerful for detecting scale increases, and $\hat{S}^{(2-)}_t$ is more powerful for detecting scale decreases. Therefore, checking which charting statistics among $\hat{S}^{(1+)}_t$, $\hat{S}^{(1-)}_t$, $\hat{S}^{(2+)}_t$, and $\hat{S}^{(2-)}_t$ have exceeded the control limit $h$ when the alarm is triggered can identify what kind of distributional changes have caused the alarm. This acts as a built-in post-signal diagnostic function, which is another appealing feature of our method.

\section{Simulation Studies}
\subsection{The proposed adaptive CUSUM chart versus the CPD charts}
In this section, we report several simulation studies to evaluate the performance of our proposed self-starting adaptive CUSUM chart for detecting different distributional changes. In particular, we compare our proposed control chart with some CPD charts, since they also do not involve any tuning parameter or require significant amount of reference data. In Ross and Adams (2012),  two CPD charts for detecting arbitrary distributional changes were developed, one is based on the Kolmogorov-Smirnov (KS) test statistic and the other on the Cramer-von-Mises (CvM) test statistic. In their conclusions,  they recommended using the CvM CPD chart, since it is usually better than the one based on the KS test statistic. In Ross, Tasoulis and Adams (2011), another CPD chart based on the Lepage test statistic was proposed. Although technically the Lepage CPD chart is only for location and scale changes, it seems to be very powerful for other situations as well.  Therefore, we include the CvM CPD chart and the Lepage CPD chart in our comparison.

To study how $d$ (the number of categories) affects the performance of our proposed control chart, we consider four choices of $d$,  $d=10$, 20, 30, and 40. Based on the simulation study conducted in Section \ref{sec:selfstart}, a warm-up period of $20$ observations can ensure good $ARL_0$ performance of our proposed self-starting control chart for those choices of $d$. For the CvM CPD chart and the Lepage CPD chart, a warm-up period of 20 observations is also recommended in Ross and Adams (2012) and Ross, Tasoulis and Adams (2011). Therefore, for all the three charts, we start monitoring only after the first 20 observations have been received. Following the simulation settings considered in Ross and Adams (2012), we  compare the performance of our proposed control chart along with the CvM CPD chart and the Lepage CPD chart for detecting location changes, scale changes and more general distributional changes.

\subsection*{Location changes}
For location changes, three different in-control distributions are considered: the standard normal, $N(0,1)$; the $t$ distribution with 2.5 degrees of freedom, $t(2.5)$; and the lognormal distribution with parameters $\mu=1$ and $\sigma=0.5$, $LN(1,0.5)$. For $t(2.5)$ and $LN(1,0.5)$, we also standardize the data so that the in-control distribution has mean 0 and standard deviation 1. We denote the resulting distributions by $t(2.5)/\sqrt{5}$ and $(LN(1,0.5)-3)/1.6$, respectively. To simulate location changes, we add a constant $\delta$ $\in \{0.25,0.5,0.75,1,1.5,2\}$ to the observations collected after the change-point $\tau$. Two choices of $\tau$ are considered: $\tau=50$ or $300$. The average time taken to detect the change (denoted by $ARL_1$) from 10,000 simulations is then recorded for each chart. Table \ref{tab:loc} shows the $ARL_1$ of all the three control charts
along with their corresponding standard errors (in the parentheses) under different settings.

\begin{table}[!htbp]
\centering
\caption{The simulated $ARL_1$ for our proposed self-starting adaptive CUSUM chart with different choices of $d$, the Lepage CPD chart and the CvM CPD chart for detecting location shifts.\label{tab:loc}}
\begin{tabular}{|r|r||rrrr||r|r|}
  \hline
\multicolumn{8}{|c|}{$N(0,1)+\delta$}\\
\hline
 & &\multicolumn{4}{|c||}{Proposed} &  & \\
\cline{3-6}
$\tau$ & $\delta$   & $d=10$ & $d=20$ & $d=30$ & $d=40$ & Lepage & CvM \\
  \hline
\hline
& 0.25 & 395.71 (4.54) & 381.98(4.52) & 373.24(4.35) & 369.53(4.36) & 436.73(4.77) & 382.95(4.63) \\
&0.50 & 179.36(3.19) & 158.55(2.95) & 151.34(2.91) & 143.85(2.76) & 232.36(3.62) & 157.97(2.91) \\
& 0.75 & 46.34(1.12) & 40.12(0.91) & 38.13(0.84) & 36.06(0.67) & 62.96(1.32) & 37.44(0.97) \\
50 & 1.00 & 17.02(0.18) & 16.78(0.14) & 17.43(0.13) & 17.82(0.12) & 20.04(0.23) & 14.85(0.14) \\
& 1.50 & 8.47(0.04) & 8.89(0.04) & 9.22(0.04) & 9.45(0.04) & 6.89(0.05) & 6.64(0.04) \\
& 2.00 & 6.08(0.02) & 6.19(0.02) & 6.41(0.02) & 6.54(0.03) & 3.67(0.02) & 4.32(0.02) \\
\hline
& 0.25 & 205.69(2.88) & 172.74(2.43) & 167.83(2.30) & 159.45(2.20) & 227.90(2.81) & 164.03(2.10) \\
& 0.50 & 40.71(0.36) & 37.77(0.29) & 37.35(0.28) & 36.57(0.26) & 49.63(0.43) & 38.29(0.32) \\
& 0.75 & 19.25(0.11) & 18.91(0.11) & 19.20(0.10) & 19.37(0.10) & 20.75(0.15) & 17.90(0.12) \\
300 & 1.00 & 12.37(0.06) & 12.43(0.06) & 12.81(0.06) & 12.91(0.06) & 11.61(0.08) & 10.78(0.06) \\
& 1.50 & 7.29(0.03) & 7.26(0.03) & 7.43(0.03) & 7.48(0.03) & 5.18(0.03) & 5.71(0.03) \\
& 2.00 & 5.41(0.02) & 5.21(0.02) & 5.22(0.02) & 5.24(0.02) & 3.06(0.02) & 3.90(0.02) \\
 \hline
  \hline
 \multicolumn{8}{|c|}{$t(2.5)/\sqrt{5}+\delta$}\\
\hline
 & &\multicolumn{4}{|c||}{Proposed} &  & \\
\cline{3-6}
$\tau$ & $\delta$   & $d=10$ & $d=20$ & $d=30$ & $d=40$ & Lepage & CvM \\
  \hline
\hline
& 0.25 & 262.22(3.94) & 256.42(3.92) & 244.59(3.80) & 239.02(3.63) & 304.36(4.11) & 194.13(3.26) \\
& 0.50 & 32.60(0.79) & 34.67(0.78) & 33.42(0.62) & 34.60(0.76) & 38.23(0.65) & 20.81(0.42) \\
& 0.75 & 11.85(0.08) & 13.20(0.08) & 14.07(0.08) & 14.83(0.09) & 11.94(0.10) & 8.58(0.06) \\
50 & 1.00 & 7.96(0.04) & 8.90(0.04) & 9.66(0.04) & 10.12(0.05) & 6.48(0.05) & 5.63(0.03) \\
& 1.50 & 5.45(0.02) & 5.95(0.02) & 6.40(0.03) & 6.69(0.03) & 3.23(0.02) & 3.77(0.01) \\
& 2.00 & 4.70(0.01) & 4.90(0.02) & 5.31(0.02) & 5.44(0.02) & 2.41(0.01) & 3.19(0.01) \\
   \hline
& 0.25 & 63.18(0.77) & 62.30(0.73) & 61.65(0.69) & 60.74(0.62) & 73.87(0.67) & 46.57(0.41) \\
& 0.50 & 16.01(0.09) & 17.28(0.09) & 18.05(0.10) & 18.56(0.10) & 16.86(0.11) & 13.10(0.08) \\
& 0.75 & 8.95(0.04) & 9.93(0.04) & 10.62(0.05) & 11.03(0.05) & 7.51(0.04) & 7.07(0.04) \\
300 & 1.00 & 6.42(0.03) & 7.00(0.03) & 7.47(0.03) & 7.82(0.03) & 4.53(0.02) & 4.95(0.02) \\
& 1.50 & 4.70(0.01) & 4.62(0.02) & 4.87(0.02) & 5.09(0.02) & 2.58(0.01) & 3.40(0.01) \\
& 2.00 & 4.23(0.01) & 3.82(0.01) & 3.86(0.01) & 3.96(0.01) & 2.05(0.01) & 3.00(0.01) \\
   \hline
  \hline
 \multicolumn{8}{|c|}{$(LN(1,0.5)-3)/1.6+\delta$}\\
\hline
 & &\multicolumn{4}{|c||}{Proposed} &  & \\
\cline{3-6}
$\tau$ & $\delta$   & $d=10$ & $d=20$ & $d=30$ & $d=40$ & Lepage & CvM \\
  \hline
\hline
& 0.25 & 330.66(4.36) & 295.84(4.21) & 282.71(3.99) & 278.45(3.96) & 412.16(4.82) & 376.09(4.70) \\
& 0.50 & 71.74(1.91) & 54.20(1.19) & 50.64(0.99) & 51.14(0.97) & 94.96(1.74) & 109.15(2.43) \\
& 0.75 & 19.37(0.19) & 20.56(0.12) & 21.44(0.11) & 22.49(0.11) & 27.33(0.21) & 23.07(0.54) \\
50 & 1.00 & 12.72(0.06) & 14.13(0.06) & 15.09(0.06) & 15.90(0.07) & 15.69(0.09) & 10.66(0.08) \\
& 1.50 & 7.98(0.03) & 9.01(0.03) & 9.77(0.04) & 10.26(0.04) & 7.71(0.04) & 5.47(0.02) \\
& 2.00 & 6.07(0.02) & 6.85(0.02) & 7.39(0.03) & 7.73(0.03) & 4.35(0.02) & 4.00(0.01) \\
   \hline
 & 0.25 & 101.39(1.37) & 84.21(0.99) & 78.17(0.80) & 75.66(0.72) & 144.45(1.45) & 122.44(1.46) \\
 & 0.50 & 24.91(0.13) & 25.52(0.12) & 26.33(0.12) & 26.89(0.12) & 37.61(0.19) & 26.90(0.17) \\
 & 0.75 & 14.53(0.06) & 15.73(0.06) & 16.56(0.06) & 17.22(0.06) & 19.41(0.09) & 13.17(0.07) \\
 300 & 1.00 & 10.49(0.04) & 11.60(0.04) & 12.29(0.04) & 12.83(0.04) & 11.99(0.06) & 8.28(0.03) \\
& 1.50 & 6.89(0.02) & 7.62(0.02) & 8.12(0.03) & 8.48(0.03) & 5.57(0.03) & 4.80(0.01) \\
& 2.00 & 5.22(0.02) & 5.67(0.02) & 6.00(0.02) & 6.29(0.02) & 3.31(0.01) & 3.57(0.01) \\
   \hline
\end{tabular}
\end{table}

As we can see from Table \ref{tab:loc}, the choice of $d$ affects the $ARL_1$ of the proposed CUSUM chart. In general, our CUSUM charts with  larger $d$ have better $ARL_1$ than those with smaller $d$ for detecting small location shifts, and vice versa for detecting large location shifts. This can be explained by the following. On one hand, our charting statistic with larger $d$ is usually more sensitive to the location changes, since it monitors the location changes in $d$ categories. Therefore, for small location shifts, our CUSUM charts with larger $d$ are more powerful. On the other hand, our charting statistic with larger $d$ requires more observations in total to build up the evidence for location changes. Therefore, for large location shifts, it takes our CUSUM charts with larger $d$ longer time to detect those changes. Considering the performance for detecting both small and large location shifts, we recommend using $d=20$ in our proposed CUSUM chart.

Now we compare our proposed CUSUM chart with the two CPD charts. Between the two CPD charts, the CvM CPD chart is generally better than the Lepage CPD chart. For small location shifts, our proposed CUSUM chart is always better than the Lepage CPD chart. Comparing with the CvM CPD chart, the performance of our CUSUM chart is similar in the normal distribution, worse in the $t$ distribution, and better in the lognormal distribution. For large location shifts, the two CPD charts are generally better than our CUSUM chart. This is because the two CPD charts are based on the ranks of the observations, while our CUSUM chart is constructed through the categorization of the observations. When the process experiences large shifts, most of the observations will have large ranks which can quickly drive the charting statistics of the two CPD charts to exceed their respective control limits. However, this ranking information will not be completely preserved through data categorization, therefore our CUSUM chart will not react as quickly as those two CPD charts to large location shifts.

\subsection*{Scale changes}
For scale changes, we also consider the three in-control distributions: $N(0,1)$, $t(2.5)/\sqrt{5}$ and $(LN(1,0.5)-3)/1.6$. To simulate scale changes, we multiply a constant $\delta$ $\in \{1.5, 2, 3,0.5, 0.33, 0.2\}$ to the observations collected after the change-point $\tau$.  Again $\tau=50$ or $300$. The first three choices of $\delta$ indicate an increase in scale, while the last three choices indicate a decrease in scale. Table \ref{tab:scal} shows the $ARL_1$ of all the three control charts
along with their corresponding standard errors (in the parentheses) from 10,000 simulations under different settings.

\begin{table}[!htbp]
\centering
\caption{The simulated $ARL_1$ for our proposed self-starting adaptive CUSUM chart with different choices of $d$, the Lepage CPD chart and the CvM CPD chart for detecting scale changes.\label{tab:scal}}
\begin{tabular}{|r|r||rrrr||r|r|}
  \hline
   \multicolumn{8}{|c|}{$N(0,1) \times \delta$}\\
\hline
 & &\multicolumn{4}{|c||}{Proposed} &  & \\
\cline{3-6}
$\tau$ & $\delta$   & $d=10$ & $d=20$ & $d=30$ & $d=40$ & Lepage & CvM \\
  \hline
  \hline
& 1.50 & 175.01(3.05) & 145.11(2.68) & 136.53(2.53) & 125.93(2.42) & 149.53(2.62) & 314.07(4.05) \\
& 2.00 & 31.32(0.62) & 27.83(0.53) & 25.79(0.45) & 23.07(0.41) & 26.89(0.58) & 202.42(3.21) \\
& 3.00 & 11.34(0.07) & 10.97(0.07) & 10.46(0.06) & 9.90(0.06) & 8.48(0.07) & 61.37(1.07) \\
50 & 0.50 & 36.99(0.87) & 33.39(0.60) & 33.07(0.44) & 33.73(0.49) & 62.46(1.28) & 562.99(5.68) \\
& 0.33 & 13.74(0.06) & 15.25(0.07) & 16.25(0.07) & 16.97(0.07) & 19.93(0.10) & 192.20(3.00) \\
& 0.20 & 9.47(0.04) & 10.59(0.04) & 11.41(0.04) & 11.98(0.05) & 13.72(0.03) & 44.90(0.36) \\
   \hline
& 1.50 & 41.04(0.37) & 34.17(0.28) & 32.25(0.25) & 31.17(0.24) & 34.57(0.33) & 131.91(1.66) \\
& 2.00 & 16.75(0.11) & 14.62(0.09) & 13.97(0.08) & 13.57(0.08) & 13.16(0.10) & 46.85(0.40) \\
 & 3.00 & 9.28(0.05) & 8.23(0.04) & 7.86(0.04) & 7.57(0.04) & 6.56(0.04) & 22.24(0.16) \\
300 & 0.50 & 18.91(0.08) & 19.93(0.08) & 20.84(0.08) & 21.40(0.09) & 30.90(0.11) & 102.58(0.43) \\
& 0.33 & 10.71(0.04) & 11.83(0.04) & 12.55(0.04) & 13.09(0.04) & 17.63(0.04) & 42.19(0.10) \\
& 0.20 & 7.42(0.02) & 8.18(0.03) & 8.76(0.03) & 9.11(0.03) & 13.49(0.02) & 25.31(0.04) \\
   \hline
  \hline
 \multicolumn{8}{|c|}{$t(2.5)/\sqrt{5} \times \delta$}\\
\hline
 & &\multicolumn{4}{|c||}{Proposed} &  & \\
\cline{3-6}
$\tau$ & $\delta$   & $d=10$ & $d=20$ & $d=30$ & $d=40$ & Lepage & CvM \\
  \hline
  \hline
& 1.50 & 266.73(3.71) & 257.87(3.66) & 250.77(3.66) & 243.65(3.60) & 252.97(3.66) & 359.59(4.34) \\
& 2.00 & 88.86(2.01) & 79.27(1.85) & 74.22(1.69) & 67.69(1.52) & 84.17(1.80) & 260.92(3.79) \\
 & 3.00 & 17.65(0.26) & 17.58(0.16) & 17.33(0.16) & 16.83(0.19) & 15.47(0.19) & 120.56(2.25) \\
50 & 0.50 & 101.00(2.33) & 87.01(1.97) & 81.13(1.86) & 79.67(1.81) & 141.32(2.58) & 613.67(5.68) \\
& 0.33  & 19.14(0.16) & 20.19(0.14) & 21.29(0.17) & 22.03(0.16) & 28.61(0.20) & 364.21(4.59) \\
& 0.20 & 11.17(0.05) & 12.24(0.05) & 13.00(0.06) & 13.66(0.06) & 16.36(0.05) & 73.66(1.04) \\
   \hline
& 1.50 & 71.40(0.89) & 65.83(0.84) & 62.91(0.72) & 61.60(0.67) & 64.71(0.74) & 179.23(2.26) \\
& 2.00 & 23.52(0.17) & 23.03(0.16) & 22.96(0.15) & 22.98(0.15) & 21.63(0.18) & 68.29(0.63) \\
 & 3.00 & 11.66(0.06) & 11.20(0.06) & 11.35(0.06) & 11.57(0.06) & 9.28(0.07) & 29.33(0.22) \\
300 & 0.50 & 27.14(0.17) & 27.45(0.16) & 27.97(0.16) & 28.47(0.16) & 45.51(0.21) & 162.65(0.93) \\
& 0.33 & 13.27(0.06) & 14.17(0.06) & 14.75(0.06) & 15.34(0.06) & 22.94(0.07) & 56.25(0.18) \\
& 0.20 & 8.44(0.03) & 9.23(0.03) & 9.86(0.04) & 10.20(0.04) & 15.63(0.03) & 30.98(0.07) \\
   \hline
  \hline
 \multicolumn{8}{|c|}{$(LN(1,0.5)-3)/1.6 \times \delta$}\\
\hline
 & &\multicolumn{4}{|c||}{Proposed} &  & \\
\cline{3-6}
$\tau$ & $\delta$   & $d=10$ & $d=20$ & $d=30$ & $d=40$ & Lepage & CvM \\
  \hline
  \hline
& 1.50 & 118.98(2.34) & 95.71(2.03) & 88.54(1.93) & 78.25(1.71) & 98.82(1.95) & 264.31(3.74) \\
&  2.00 & 22.43(0.35) & 20.01(0.22) & 18.81(0.24) & 17.43(0.19) & 17.45(0.23) & 127.10(2.42) \\
 & 3.00 & 10.30(0.06) & 9.85(0.06) & 9.60(0.06) & 9.20(0.05) & 7.45(0.06) & 33.83(0.39) \\
50 &  0.50 & 28.75(0.57) & 27.65(0.39) & 28.31(0.41) & 29.93(0.44) & 43.64(0.74) & 434.44(5.24) \\
& 0.33 & 13.51(0.06) & 14.92(0.07) & 15.88(0.07) & 16.59(0.07) & 18.72(0.06) & 109.94(2.07) \\
& 0.20 & 10.24(0.04) & 11.44(0.05) & 12.26(0.05) & 12.84(0.05) & 13.89(0.03) & 34.77(0.24) \\
   \hline
& 1.50 & 31.32(0.25) & 25.76(0.19) & 23.94(0.17) & 22.85(0.16) & 24.76(0.22) & 85.38(0.92) \\
& 2.00 & 14.30(0.09) & 12.34(0.07) & 11.70(0.07) & 11.18(0.06) & 10.67(0.08) & 34.16(0.27) \\
 & 3.00 & 8.67(0.04) & 7.51(0.04) & 7.13(0.03) & 6.88(0.03) & 5.88(0.04) & 18.08(0.13) \\
300 & 0.50 & 17.42(0.08) & 18.54(0.08) & 19.46(0.08) & 19.99(0.08) & 27.37(0.08) & 65.02(0.25) \\
&  0.33 & 10.79(0.04) & 11.69(0.04) & 12.43(0.05) & 12.85(0.05) & 17.56(0.04) & 31.45(0.08) \\
&  0.20 & 8.68(0.03) & 9.58(0.04) & 10.11(0.04) & 10.49(0.04) & 14.35(0.02) & 20.80(0.04) \\
   \hline
\end{tabular}
\end{table}

As seen from Table \ref{tab:scal}, the performance of our proposed CUSUM chart also depends on the choice of $d$. In general, our CUSUM charts with  larger $d$ have better $ARL_1$ than those with smaller $d$ for detecting scale increases, and vice versa for detecting scale decreases. Based on the performance for detecting both scale increases and decreases, we again recommend using $d=20$ in our proposed CUSUM chart.

Between the two CPD chart, the Lepage CPD chart is much better than the CvM CPD for detecting scale changes. Comparing with the Lepage CPD chart, the performance of our CUSUM chart is similar for detecting scale increases, and much better for detecting scale decreases.

\subsection*{More general changes}
For more general distributional changes, we follow the settings considered in Ross and Adams (2012), and the eight types of distributional changes considered in their paper are listed in Table \ref{tab:change}. Again the change occurs after the change-point $\tau=50$ or $300$. Table \ref{tab:gen} shows the $ARL_1$ of all the three control charts
along with their corresponding standard errors (in the parentheses) from 10,000 simulations under the eight different distributional changes.

\begin{table}[!htbp]
\centering
\caption{The type of more general changes considered in the simulations.\label{tab:change}}
\begin{tabular}{|c|c|}
  \hline
 Change Type &   \\
  \hline
  \hline
1 & Exp(1) $\rightarrow$ Exp(3)\\
 2 & Exp(3) $\rightarrow$ Exp(1)\\
 3 &  Gamma(2,2) $\rightarrow$ Gamma(3,2)\\
 4 & Gamma(3,2) $\rightarrow$ Gamma(2,2) \\
 5 & Weibull(1) $\rightarrow$ Weibull(3)\\
6 & Weibull(3) $\rightarrow$ Weibull(1)\\
7 & Uniform(0,1) $\rightarrow$ Beta(5,5) \\
8 & Beta(5,5) $\rightarrow$ Uniform(0,1) \\
   \hline
\end{tabular}
\end{table}

\begin{table}[!htbp]
\centering
\caption{The simulated $ARL_1$ for our proposed self-starting adaptive CUSUM chart with different choices of $d$, the Lepage CPD chart and the CvM CPD chart for detecting general distributional changes.\label{tab:gen}}
\begin{tabular}{|r|c||rrrr||r|r|}
  \hline
 & Change &\multicolumn{4}{|c||}{Proposed} &  & \\
\cline{3-6}
$\tau$ & type   & $d=10$ & $d=20$ & $d=30$ & $d=40$ & Lepage & CvM \\
  \hline
  \hline
& 1 & 18.88(0.37) & 19.30(0.18) & 20.00(0.12) & 20.77(0.19) & 25.57(0.28) & 18.16(0.26) \\
&  2 & 16.19(0.19) & 15.34(0.14) & 15.20(0.11) & 15.35(0.11) & 14.67(0.21) & 14.44(0.13) \\
&  3 & 65.01(1.68) & 53.07(1.27) & 55.03(1.37) & 50.15(1.16) & 89.08(1.80) & 54.22(1.23) \\
&  4 & 61.06(1.56) & 51.51(1.25) & 46.40(1.06) & 46.65(1.18) & 82.00(1.86) & 49.80(1.16) \\
50 &  5 & 17.38(0.09) & 18.98(0.09) & 20.14(0.09) & 21.09(0.10) & 23.26(0.11) & 182.70(3.10) \\
& 6 & 12.74(0.09) & 12.13(0.08) & 11.83(0.08) & 11.10(0.07) & 9.79(0.09) & 54.05(0.93) \\
& 7 & 18.76(0.20) & 19.87(0.11) & 21.08(0.11) & 21.82(0.11) & 29.12(0.33) & 392.65(4.79) \\
& 8 & 15.97(0.14) & 14.62(0.11) & 14.07(0.10) & 13.12(0.09) & 12.49(0.13) & 119.02(2.19) \\
   \hline
& 1 & 13.55(0.06) & 14.56(0.06) & 15.40(0.06) & 15.89(0.06) & 16.82(0.09) & 11.81(0.06) \\
& 2 & 11.72(0.06) & 10.88(0.06) & 10.76(0.06) & 10.61(0.05) & 8.48(0.06) & 10.82(0.07) \\
& 3 & 22.43(0.13) & 21.92(0.12) & 22.23(0.12) & 22.67(0.12) & 27.33(0.19) & 20.90(0.14) \\
& 4 & 21.58(0.14) & 20.30(0.12) & 20.64(0.12) & 20.48(0.12) & 21.72(0.17) & 20.31(0.15) \\
300 & 5 & 14.38(0.05) & 15.71(0.06) & 16.61(0.06) & 17.17(0.06) & 20.69(0.05) & 37.74(0.13) \\
& 6 & 9.98(0.05) & 8.93(0.05) & 8.52(0.04) & 8.32(0.04) & 7.16(0.05) & 21.75(0.16) \\
& 7 & 13.61(0.05) & 14.80(0.05) & 15.60(0.06) & 16.19(0.06) & 22.08(0.06) & 60.33(0.18) \\
& 8 & 11.69(0.06) & 10.25(0.05) & 9.96(0.05) & 9.63(0.05) & 8.52(0.06) & 30.55(0.23) \\
   \hline
\end{tabular}
\end{table}

As we can see from Table \ref{tab:gen}, different choices of $d$ make slight differences in $ARL_1$ for our proposed CUSUM chart. Our recommendation $d=20$ from the previous simulation studies also seems to work well in all the settings considered here. Between the two CPD charts, there is no clear winner: the CvM CPD chart works better in change types 1, 3 and 4, while the Lepage CPD chart works better in change types 5, 6, 7 and 8. Among all eight change types, we can see that, if our proposed CUSUM chart is not the best, it is very close to the best.

In summary, based on the three simulation studies presented above for detecting different types of distributional changes, our proposed CUSUM chart is the best in overall performance comparing with the other two CPD charts. Coupling with its computational advantage over the two CPD charts, our proposed CUSUM chart proves to be a flexible and efficient monitoring tool.

\subsection{The proposed adaptive CUSUM chart versus other possible nonparametric adaptive CUSUM charts}
In Section \ref{sec:CUSUM}, before we get to the CUSUM statistic $S_t$ in (\ref{eqn:CUSUM_loc_scal}), we also describe several other possible CUSUM statistics based on the categorized data. For example,  $S^{(0)}_t$ defined in (\ref{eqn:CUSUM0}) directly uses the categorized data $\mbs{Y}^{(1)}_t$, $S^{(1)}_t$ in (\ref{eqn:CUSUM_loc}) makes use of the left-to-right ordering of the data, and $S^{(2)}_t$ in (\ref{eqn:CUSUM_scal}) incorporates the center-outward ordering of the data. Similar to the approaches presented in Sections 2.2-2.4, based on the CUSUM statistics $S^{(0)}_t$, $S^{(1)}_t$ and $S^{(2)}_t$, we can also develop their self-starting adaptive CUSUM charts, and their corresponding charting statistics are denoted by $\hat{S}^{(0)}_t$, $\hat{S}^{(1)}_t$ and $\hat{S}^{(2)}_t$, respectively. In this section,  we compare those control charts with our proposed self-starting adaptive CUSUM chart based on the charting statistic $\hat{S}_t$ in (\ref{eqn:adaptiveCUSUM_loc_scal}). This is to demonstrate the reason described in Section \ref{sec:CUSUM} when we choose $S_t$ as our CUSUM statistic. The simulation settings we consider in this section are the same as those in the previous section. Tables \ref{tab:loc1}-\ref{tab:gen1} summarize the $ARL_1$ of the four control charts along with their corresponding standard errors (in the parentheses) from 10,000 simulations under those settings. In all four control charts, we set $d=20$.

\begin{table}[!htbp]
\centering
\caption{The simulated $ARL_1$ for the self-starting adaptive CUSUM chart based on  $\hat{S}_t^{(0)}$, $\hat{S}_t^{(1)}$, $\hat{S}_t^{(2)}$ and  $\hat{S}_t$ for detecting location shifts.\label{tab:loc1}}
\begin{tabular}{|r|r||r|rr|r|}
  \hline
   \multicolumn{6}{|c|}{$N(0,1) + \delta$}\\
\hline
$\tau$ & $\delta$   &  $\hat{S}_t^{(0)}$ & $\hat{S}_t^{(1)}$ & $\hat{S}_t^{(2)}$ & $\hat{S}_t$ \\
  \hline
  \hline
& 0.25 & 447.58(4.79) & 359.45(4.40) & 491.15(4.88) & 381.98(4.52) \\
&  0.50 & 309.99(4.30) & 126.72(2.67) & 479.39(4.84) & 158.55(2.95) \\
& 0.75 & 147.60(3.03) & 30.48(0.67) & 458.07(4.83) & 40.12(0.91) \\
50 & 1.00 & 49.77(1.51) & 13.93(0.10) & 384.61(4.64) & 16.78(0.14) \\
  & 1.50 & 10.20(0.07) & 7.61(0.03) & 149.65(3.01) & 8.89(0.04) \\
  & 2.00 & 6.52(0.03) & 5.40(0.02) & 25.81(0.98) & 6.19(0.02) \\
   \hline
& 0.25 & 308.95(3.93) & 144.36(2.11) & 473.90(4.85) & 172.74(2.43) \\
&  0.50 & 84.56(1.36) & 32.66(0.26) & 413.94(4.66) & 37.77(0.29) \\
&  0.75 & 26.93(0.26) & 16.43(0.10) & 246.76(3.89) & 18.91(0.11) \\
300 &  1.00 & 14.77(0.09) & 10.85(0.05) & 69.28(1.82) & 12.43(0.06) \\
 &  1.50 & 7.69(0.03) & 6.46(0.03) & 11.32(0.07) & 7.26(0.03) \\
&  2.00 & 5.11(0.02) & 4.71(0.02) & 6.36(0.03) & 5.21(0.02) \\
   \hline
     \hline
 \multicolumn{6}{|c|}{$t(2.5)/\sqrt{5}+\delta$}\\
\hline
$\tau$ & $\delta$   &  $\hat{S}_t^{(0)}$ & $\hat{S}_t^{(1)}$ & $\hat{S}_t^{(2)}$ & $\hat{S}_t$ \\
  \hline
  \hline
& 0.25 & 344.73(4.52) & 216.33(3.61) & 483.24(4.87) & 256.42(3.92) \\
&  0.50 & 98.96(2.41) & 27.00(0.63) & 379.85(4.36) & 34.67(0.78) \\
&  0.75 & 17.16(0.45) & 11.23(0.07) & 215.96(3.51) & 13.20(0.08) \\
50 &  1.00 & 9.16(0.11) & 7.68(0.04) & 83.47(2.09) & 8.90(0.04) \\
&  1.50 & 5.82(0.02) & 5.19(0.02) & 12.20(0.34) & 5.95(0.02) \\
&  2.00 & 4.71(0.02) & 4.29(0.02) & 6.82(0.06) & 4.90(0.02) \\
  \hline
& 0.25 & 128.71(2.10) & 51.20(0.57) & 448.46(4.73) & 62.30(0.73) \\
&  0.50 & 20.10(0.15) & 15.14(0.09) & 127.25(2.59) & 17.28(0.09) \\
&  0.75 & 10.30(0.05) & 8.84(0.04) & 18.35(0.22) & 9.93(0.04) \\
300 &  1.00 & 7.04(0.03) & 6.25(0.03) & 10.05(0.05) & 7.00(0.03) \\
&  1.50 & 4.41(0.02) & 4.25(0.02) & 5.65(0.02) & 4.62(0.02) \\
&  2.00 & 3.46(0.01) & 3.64(0.01) & 4.25(0.02) & 3.82(0.01) \\
   \hline
   \hline
 \multicolumn{6}{|c|}{$(LN(1,0.5)-3)/1.6+\delta$}\\
\hline
$\tau$ & $\delta$   &  $\hat{S}_t^{(0)}$ & $\hat{S}_t^{(1)}$ & $\hat{S}_t^{(2)}$ & $\hat{S}_t$ \\
  \hline
  \hline
& 0.25 & 442.91(4.87) & 299.41(4.25) & 448.83(4.81) & 295.84(4.21) \\
&  0.50 & 253.46(4.11) & 49.25(1.24) & 335.58(4.25) & 54.20(1.19) \\
&  0.75 & 79.66(2.13) & 17.53(0.11) & 241.27(3.34) & 20.56(0.12) \\
50 &  1.00 & 23.01(0.72) & 12.00(0.05) & 188.16(2.48) & 14.13(0.06) \\
&  1.50 & 9.38(0.04) & 7.69(0.03) & 96.86(1.54) & 9.01(0.03) \\
&  2.00 & 6.80(0.03) & 5.88(0.02) & 27.65(0.59) & 6.85(0.02) \\
   \hline
& 0.25 & 268.36(3.59) & 78.77(1.04) & 336.51(3.98) & 84.21(0.99) \\
&  0.50 & 48.67(0.61) & 22.41(0.11) & 214.29(2.08) & 25.52(0.12) \\
&  0.75 & 19.36(0.11) & 13.61(0.05) & 204.82(1.49) & 15.73(0.06) \\
300 &  1.00 & 12.49(0.05) & 10.08(0.04) & 165.66(1.49) & 11.60(0.04) \\
&  1.50 & 7.64(0.03) & 6.67(0.02) & 13.61(0.09) & 7.62(0.02) \\
&  2.00 & 5.56(0.02) & 5.01(0.02) & 7.70(0.03) & 5.67(0.02) \\
   \hline
\end{tabular}
\end{table}

\begin{table}[!htbp]
\centering
\caption{The simulated $ARL_1$ for the self-starting adaptive CUSUM chart based on  $\hat{S}_t^{(0)}$, $\hat{S}_t^{(1)}$, $\hat{S}_t^{(2)}$ and  $\hat{S}_t$ for detecting scale changes.\label{tab:scal1}}
\begin{tabular}{|r|r||r|rr|r|}
  \hline
   \multicolumn{6}{|c|}{$N(0,1)
   \times \delta$}\\
\hline
$\tau$ & $\delta$   &  $\hat{S}_t^{(0)}$ & $\hat{S}_t^{(1)}$ & $\hat{S}_t^{(2)}$ & $\hat{S}_t$ \\
  \hline
  \hline
& 1.50 & 362.25(4.40) & 316.70(3.96) & 123.90(2.44) & 145.11(2.68) \\
& 2.00 & 214.70(3.54) & 188.35(3.11) & 23.02(0.46) & 27.83(0.53) \\
& 3.00  & 55.36(1.56) & 52.00(1.23) & 9.52(0.06) & 10.97(0.07) \\
50 & 0.50 & 425.12(4.97) & 142.66(2.71) & 28.53(0.57) & 33.39(0.60) \\
 & 0.33 & 224.94(4.03) & 39.88(0.27) & 13.04(0.06) & 15.25(0.07) \\
 & 0.20 & 74.07(2.07) & 29.62(0.07) & 9.07(0.04) & 10.59(0.04) \\
   \hline
& 1.50 & 127.00(1.94) & 94.94(1.27) & 30.99(0.26) & 34.17(0.28) \\
& 2.00 & 30.82(0.35) & 31.25(0.21) & 13.07(0.08) & 14.62(0.09) \\
& 3.00 & 13.46(0.08) & 16.50(0.09) & 7.35(0.04) & 8.23(0.04) \\
50 & 0.50 & 177.71(2.82) & 48.16(0.15) & 17.12(0.07) & 19.93(0.08) \\
 & 0.33 & 40.81(0.56) & 31.82(0.07) & 10.17(0.04) & 11.83(0.04) \\
 & 0.20 & 20.81(0.15) & 25.62(0.05) & 7.07(0.02) & 8.18(0.03) \\
   \hline
     \hline
 \multicolumn{6}{|c|}{$t(2.5)/\sqrt{5} \times \delta$}\\
\hline
$\tau$ & $\delta$   &  $\hat{S}_t^{(0)}$ & $\hat{S}_t^{(1)}$ & $\hat{S}_t^{(2)}$ & $\hat{S}_t$ \\
  \hline
  \hline
& 1.50 & 411.34(4.48) & 377.21(4.38) & 246.63(3.72) & 257.87(3.66) \\
& 2.00 & 299.55(4.07) & 275.83(3.91) & 63.94(1.57) & 79.27(1.85) \\
& 3.00 & 145.71(2.85) & 149.81(2.91) & 15.06(0.13) & 17.58(0.16) \\
50 & 0.50 & 473.84(5.11) & 378.01(4.87) & 69.85(1.78) & 87.01(1.97) \\
& 0.33 & 333.34(4.70) & 83.60(1.59) & 17.04(0.14) & 20.19(0.14) \\
& 0.20 & 136.31(3.05) & 35.46(0.13) & 10.51(0.05) & 12.24(0.05) \\
  \hline
& 1.50 & 229.07(3.04) & 180.72(2.39) & 58.62(0.72) & 65.83(0.84) \\
& 2.00 & 73.63(1.09) & 57.14(0.63) & 20.56(0.15) & 23.03(0.16) \\
& 3.00 & 20.62(0.15) & 22.85(0.14) & 10.12(0.05) & 11.20(0.06) \\
300 & 0.50 & 302.79(4.30) & 80.13(0.57) & 23.40(0.14) & 27.45(0.16) \\
& 0.33 & 73.34(1.47) & 40.65(0.13) & 12.21(0.05) & 14.17(0.06) \\
& 0.20 & 26.23(0.22) & 29.24(0.06) & 7.97(0.03) & 9.23(0.03) \\
   \hline
   \hline
 \multicolumn{6}{|c|}{$(LN(1,0.5)-3)/1.6 \times \delta$}\\
\hline
$\tau$ & $\delta$   &  $\hat{S}_t^{(0)}$ & $\hat{S}_t^{(1)}$ & $\hat{S}_t^{(2)}$ & $\hat{S}_t$ \\
  \hline
  \hline
& 1.50 & 303.53(4.14) & 242.17(3.54) & 92.41(2.16) & 95.71(2.03) \\
& 2.00 & 136.61(2.90) & 98.81(2.11) & 17.79(0.23) & 20.01(0.22) \\
& 3.00 & 30.82(0.97) & 30.38(0.72) & 8.70(0.05) & 9.85(0.06) \\
50 & 0.50 & 349.78(4.53) & 76.41(1.51) & 25.62(0.48) & 27.65(0.39) \\
& 0.33 & 154.09(3.36) & 33.95(0.13) & 13.27(0.07) & 14.92(0.07) \\
 & 0.20 & 52.25(1.73) & 26.63(0.08) & 10.29(0.05) & 11.44(0.05) \\
   \hline
& 1.50 & 63.78(0.98) & 51.25(0.48) & 23.95(0.19) & 25.76(0.19) \\
& 2.00 & 20.56(0.15) & 22.64(0.15) & 11.16(0.07) & 12.34(0.07) \\
& 3.00 & 11.31(0.06) & 14.07(0.08) & 6.80(0.03) & 7.51(0.04) \\
300 & 0.50 & 91.68(1.68) & 37.50(0.11) & 16.19(0.08) & 18.54(0.08) \\
& 0.33 & 26.63(0.23) & 26.92(0.06) & 10.17(0.04) & 11.69(0.04) \\
& 0.20 & 15.09(0.10) & 22.35(0.05) & 8.33(0.03) & 9.58(0.04) \\
   \hline
\end{tabular}
\end{table}

\begin{table}[!htbp]
\centering
\caption{The simulated $ARL_1$ for the self-starting adaptive CUSUM chart based on  $\hat{S}_t^{(0)}$, $\hat{S}_t^{(1)}$, $\hat{S}_t^{(2)}$ and  $\hat{S}_t$ for detecting general distributional changes.\label{tab:gen1}}
\begin{tabular}{|r|r||r|rr|r|}
  \hline
$\tau$ & Change type   &  $\hat{S}_t^{(0)}$ & $\hat{S}_t^{(1)}$ & $\hat{S}_t^{(2)}$ & $\hat{S}_t$ \\
  \hline
  \hline
 & 1 & 67.25(0.36) & 16.26(0.17) & 307.53(3.87) & 19.30(0.18) \\
 & 2 & 193.37(1.14) & 13.29(0.11) & 187.03(3.57) & 15.34(0.14) \\
 & 3 & 67.25(0.36) & 40.29(1.01) & 470.85(4.89) & 53.07(1.27) \\
 & 4 & 193.37(1.14) & 37.44(0.89) & 441.57(4.86) & 51.51(1.25) \\
50 & 5 & 67.25(0.36) & 40.43(0.38) & 17.22(0.10) & 18.98(0.09) \\
 & 6 & 193.37(1.14) & 54.56(1.40) & 10.66(0.07) & 12.13(0.08) \\
 & 7 & 67.25(0.36) & 57.22(0.83) & 16.95(0.10) & 19.87(0.11) \\
 & 8 & 193.37(1.14) & 105.21(2.25) & 12.56(0.09) & 14.62(0.11) \\
   \hline
&  1 & 17.81(0.10) & 12.61(0.05) & 256.40(2.22) & 14.56(0.06) \\
& 2 & 12.09(0.07) & 10.05(0.05) & 15.63(0.11) & 10.88(0.06) \\
& 3 & 34.29(0.36) & 19.04(0.11) & 385.35(4.43) & 21.92(0.12) \\
& 4 & 29.47(0.28) & 17.88(0.11) & 159.17(3.05) & 20.30(0.12) \\
300 & 5 & 31.77(0.37) & 29.87(0.08) & 13.69(0.05) & 15.71(0.06) \\
& 6 & 14.26(0.08) & 17.24(0.10) & 8.13(0.04) & 8.93(0.05) \\
& 7 & 76.41(1.36) & 37.38(0.09) & 12.75(0.05) & 14.80(0.05) \\
& 8 & 17.94(0.12) & 21.23(0.13) & 9.24(0.05) & 10.25(0.05) \\
\hline
\end{tabular}
\end{table}

From Tables \ref{tab:loc1}-\ref{tab:gen1}, we can see that the adaptive CUSUM chart based on $\hat{S}^{(1)}_t$ is the most efficient among the four control charts for detecting location shifts. This is due to the fact that $\hat{S}^{(1)}_t$ makes use of the left-to-right ordering of the data. Similarly, because $\hat{S}^{(2)}_t$ makes use of the center-outward ordering of the data, the adaptive CUSUM chart based on $\hat{S}^{(2)}_t$ is the most efficient for detecting scale changes. Our proposed CUSUM charting statistic $\hat{S}_t$ is simply the maximum of $\hat{S}^{(1)}_t$ and $\hat{S}^{(2)}_t$, therefore it takes advantage of the benefits of both $\hat{S}^{(1)}_t$ and $\hat{S}^{(2)}_t$ and is capable of detecting both location and scale changes in an efficient manner. In contrast, the adaptive CUSUM chart based on $\hat{S}^{(0)}_t$ performs the worst among the four control charts in most of the settings considered here. This can be explained by the fact that $\hat{S}^{(0)}_t$ is based on the categorized data $\mbs{Y}^{(1)}_t$ directly and fails to make use of the ordering information of the data. This simulation study shows the importance of preserving the ordering information of the data when designing nonparametric control charts through data categorization.

\section{Real data application}
\begin{figure}[!htpb]
\begin{center}
\begin{tabular}{c}
\includegraphics[width=3.3in,height=3.3in]{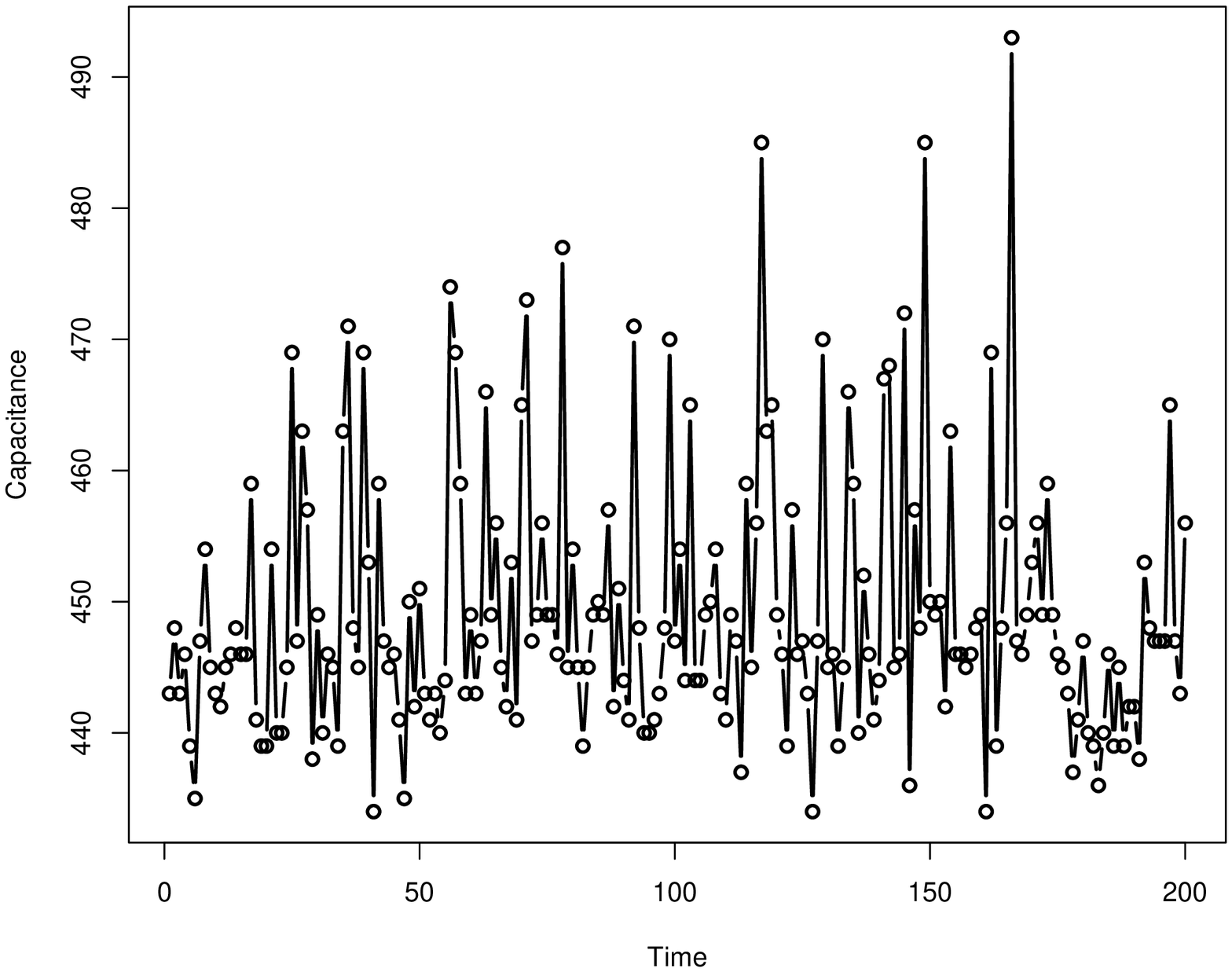}\\(a)
\end{tabular}
\begin{tabular}{c}
\includegraphics[width=3.3in,height=3.3in]{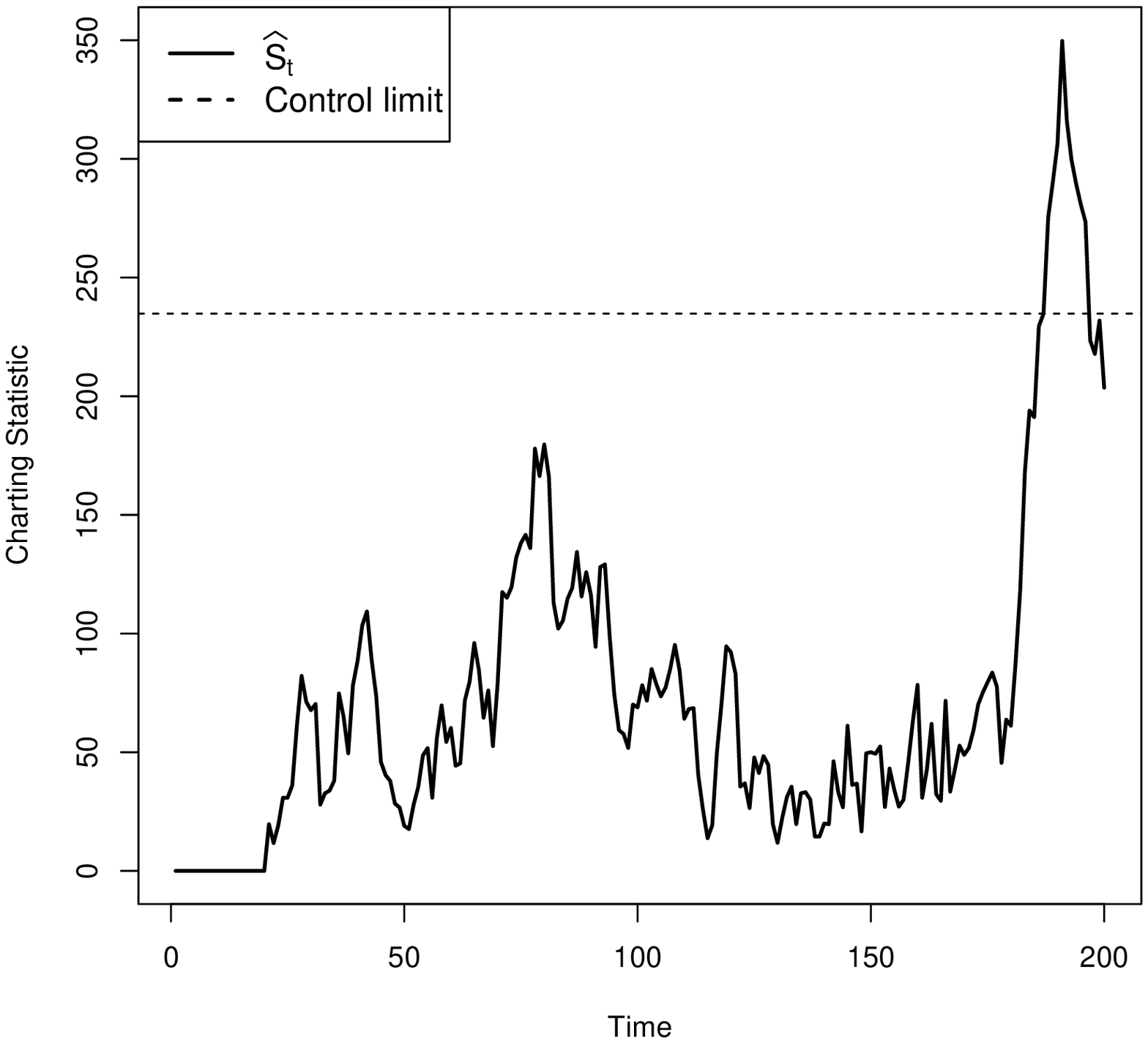}\\(b)
\end{tabular}
\caption{(a) The time series plot of the AEC data. (b) Our proposed control chart based on $\hat{S}_t$ for monitoring the AEC data.}\label{fig:data}
\end{center}
\end{figure}

In this section, we use a data set in Zou and Tsung (2010) to demonstrate the application of our proposed control chart. The data set consists of 200 observations collected from an aluminium electrolytic capacitor (AEC) manufacturing process, and each observation is the capacitance level of the AEC. Figure \ref{fig:data}(a) shows the time series plot of those 200 observations. As shown in Zou and Tsung (2010), the normality assumption does not hold for this data set, therefore some nonparametric control chart is more suitable in this application. We apply our proposed self-starting adaptive CUSUM chart to this data set. Similar to our simulation study, we set the $\text{ARL}_0$ to be 500, choose $d$ to be 20, and start monitoring after the first 20 observations.  Figure  \ref{fig:data}(b) shows the trajectory of our proposed charting statistic over the time.

\begin{figure}[!htpb]
\begin{center}
\begin{tabular}{c}
\includegraphics[width=3.3in,height=3.3in]{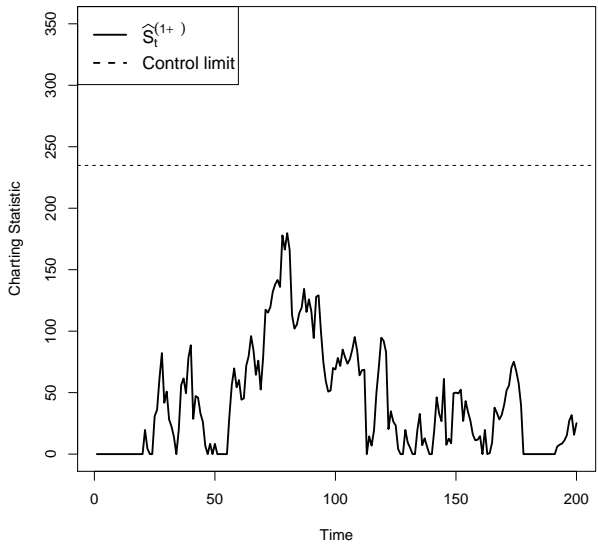} \\(a)
\end{tabular}
\begin{tabular}{c}
\includegraphics[width=3.3in,height=3.3in]{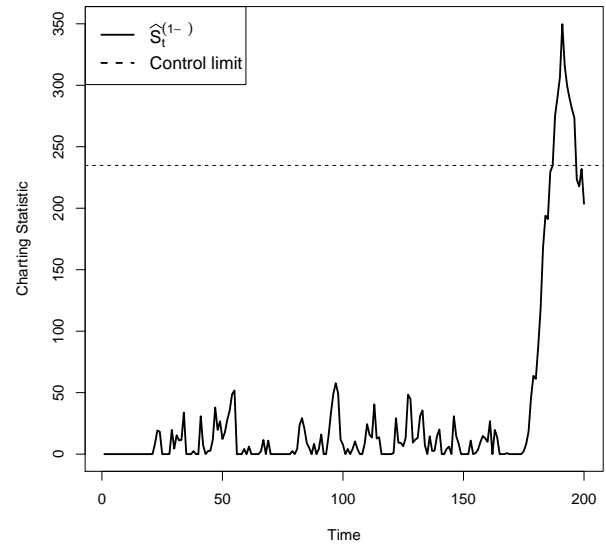} \\(b)
\end{tabular}
\begin{tabular}{c}
\includegraphics[width=3.3in,height=3.3in]{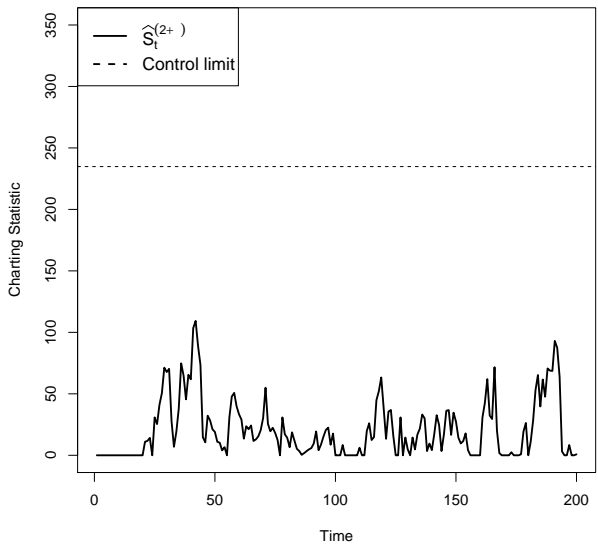} \\(c)
\end{tabular}
\begin{tabular}{c}
\includegraphics[width=3.3in,height=3.3in]{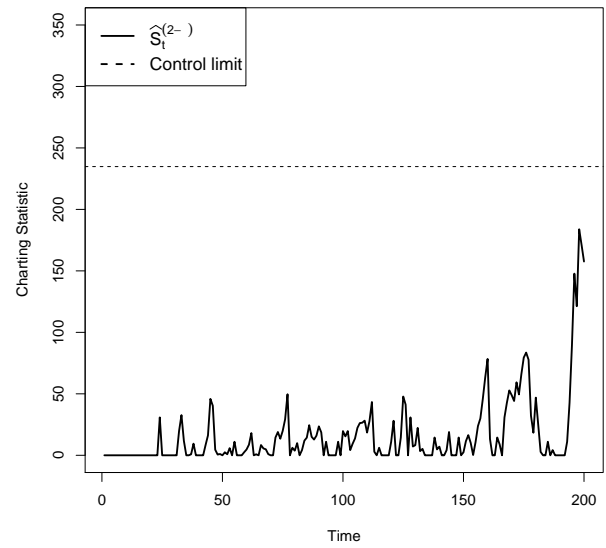} \\(d)
\end{tabular}
\caption{The control chart based on (a) $\hat{S}^{(1+)}$; (b) $\hat{S}^{(1-)}$; (c) $\hat{S}^{(2+)}$; (d) $\hat{S}^{(2-)}$ for monitoring the AEC data.}\label{fig:CUSUMs}
\end{center}
\end{figure}

As seen from Figure \ref{fig:data}(b), our proposed control chart triggers an alarm at the 188th observation. In addition to detecting the change, we are also interested in identifying what kind of distributional changes have triggered the alarm. As mentioned in Section \ref{sec:postsignal}, our proposed monitoring scheme is equivalent to monitoring $\hat{S}^{(1+)}_t$, $\hat{S}^{(1-)}_t$, $\hat{S}^{(2+)}_t$, and $\hat{S}^{(2-)}_t$ separately, and raising an alarm whenever at least one of them exceeds the control limit. Figures \ref{fig:CUSUMs}(a)-(d) show the trajectories of $\hat{S}^{(1+)}_t$, $\hat{S}^{(1-)}_t$, $\hat{S}^{(2+)}_t$, and $\hat{S}^{(2-)}_t$ over the time. From Figure \ref{fig:CUSUMs}, we can see that the alarm is mainly caused by  $\hat{S}^{(1-)}_t$. Recall that $\hat{S}^{(1+)}_t$ is more powerful for detecting positive location shifts, $\hat{S}^{(1-)}_t$ is more powerful for detecting negative location shifts, $\hat{S}^{(2+)}_t$ is more powerful for detecting scale increases, and $\hat{S}^{(2-)}_t$ is more powerful for detecting scale decreases. From the above, we can conclude that the process is experiencing a negative location shift. This seems to be consistent with what can be observed from the time series plot of the data in Figure \ref{fig:data}(a).

\section{Concluding remarks}
In this paper, we propose a nonparametric adaptive CUSUM chart for detecting arbitrary distributional changes. It is free of any tuning parameter, easy to implement and fast in computation. It does not require a large reference data set to start with due to its self-starting nature. It can also automatically identify the distributional changes once an alarm is triggered. Our simulation studies show that the overall performance of the proposed control chart is the best comparing with other existing nonparametric control charts for detecting a variety of distributional changes. All the above features make our proposed control chart very attractive to use in practice.

Although our proposed control chart is for detecting any arbitrary distributional changes, based on its construction we can easily develop other efficient nonparametric control charts if only certain types of distributional changes are of interest. For example, if we are only concerned about positive location shifts, we can build our control chart based on  $\hat{S}^{(1+)}_t$. Similarly, for negative location shifts, we can use $\hat{S}^{(1-)}_t$; for scale increases, we can use $\hat{S}^{(2+)}_t$; and for scale decreases, we can use $\hat{S}^{(2-)}_t$. If we are only interested in detecting location shifts (both positive and negative), we can use $\max(\hat{S}^{(1+)}_t,\hat{S}^{(1-)}_t)$. If we are only interested in detecting scale changes, we can use $\max(\hat{S}^{(2+)}_t,\hat{S}^{(2-)}_t)$. If scale decreases are not particularly of interest, we can use $\max(\hat{S}^{(1+)}_t,\hat{S}^{(1-)}_t, \hat{S}^{(2+)}_t)$. From the above, we can see that our proposed charting statistic also offers many possibilities to construct other efficient nonparametric control charts for detecting certain types of distributional changes. We plan to further evaluate the performance of those control charts in our future studies.

\section*{Appendix: Proof}
\begin{proof}[\textbf{Proof of Theorem \ref{thm1}}]
Based on the probability integral transformation, without loss of generality we assume that the in-control distribution of $X_t$ is the uniform distribution on (0,1). It is clear that $\hat{\mbs{Y}}^{(i)}_{t}$ follows a multinomial distribution. Note that
\begin{align*}
&P\left(X_t  \in (0, \hat{q}^{(1)}_{t,j}]\right)=E\left(\hat{q}^{(1)}_{t,j}\right)\\
=&\left(1-\frac{j(m+t)}{d}+l\right) E\left(X_{t,(l)}\right)+\left(\frac{j(m+t)}{d}-l\right)E\left(X_{t,(l+1)}\right)
\end{align*}
where $ l/(m+t) \leq j/d < (l+1)/(m+t)$.
Since the in-control distribution of $X_t$ is the uniform distribution on (0,1), the order statistics $X_{t,(l)}$ and $X_{t,(l+1)}$ follow the beta distribution beta$(l,m+t-l)$ and beta$(l+1,m+t-l-1)$, respectively. Therefore,
\[
P\left(X_t  \in (0, \hat{q}^{(1)}_{t,j}]\right)
=\left(1-\frac{j(m+t)}{d}+l\right) E\left(X_{t,(l)}\right)+\left(\frac{j(m+t)}{d}-l\right)E\left(X_{t,(l+1)}\right)=j/d.
\]
As a result,
\[
P(\hat{Y}^{(1)}_{t,j} =1)=P\left(X_t  \in (\hat{q}^{(1)}_{t,j-1}, \hat{q}^{(1)}_{t,j}]\right)=P\left(X_t  \in (0, \hat{q}^{(1)}_{t,j}]\right)-P\left(X_t  \in (0, \hat{q}^{(1)}_{t,j-1}]\right)=1/d,
\]
Similarly, we can obtain
\[
P\left(X_t  \in (0, \hat{q}^{(2)}_{t,j}]\right)
=\left(1-\frac{j(m+t)}{2d}+l\right) E\left(X_{t,(l)}\right)+\left(\frac{j(m+t)}{2d}-l\right)E\left(X_{t,(l+1)}\right)=j/(2d),
\]
where $ l/(m+t) \leq j/(2d) < (l+1)/(m+t)$,
and
\[
P(\hat{Y}^{(2)}_{t,j} =1)=P\left(X_t  \in (\hat{q}^{(2)}_{t,k-j},\hat{q}^{(2)}_{t,k-j+1}]\right)+P\left(X_t \in (  \hat{q}^{(2)}_{t,k+j-1}, \hat{q}^{(2)}_{t,k+j}]\right)=1/d.
\]
Therefore, both $\hat{\mbs{Y}}^{(1)}_{t}$ and $\hat{\mbs{Y}}^{(2)}_{t}$  follow Multi$(1;1/d,...,1/d)$, the same as $\mbs{Y}^{(1)}_{t}$ and $\mbs{Y}^{(2)}_{t}$.

To prove that the $\hat{\mbs{Y}}^{(i)}_{t}$, $i=1,2$, are independently distributed among different $t$, we notice that the sequential rank of $X_t$, i.e., the rank of $X_t$ in the set $X_{-m+1},...,X_0, X_1,...,X_{t-1},X_t$, independently follows a uniform distribution on the integers 1,2,..., $m+t$. Define $\hat{C}_{t,1}=(0,X_{t,(1)}], \, \hat{C}_{t,2}=(X_{t,(1)}, X_{t,(2)}], ..., \hat{C}_{t,m+t}=(X_{t,(m+t-1)},1)$. The above independence of the sequential ranks implies that the probabilities of $X_t$ falling in the intervals $\hat{C}_{t,1},..., \hat{C}_{t,m+t}$ are independent among different $t$. Since $\hat{A}^{(i)}_{t,j}$, $i=1,2$ and $j=1,...,d$, can be all constructed from  $\hat{C}_{t,1},..., \hat{C}_{t,m+t}$, the probabilities of $X_t$ falling in the regions $\hat{A}^{(i)}_{t,1},..., \hat{A}^{(i)}_{t,d}$ are also independent among different $t$. This proves that $\hat{\mbs{Y}}^{(i)}_{t}$, $i=1,2$, are independently distributed among different $t$.
\end{proof}
\nocite*{}

\end{document}